\documentclass[a4paper,11pt]{article}
\usepackage[dvipsnames]{xcolor}
\usepackage[utf8]{inputenc}
\usepackage[bbgreekl]{mathbbol}
\usepackage{geometry}
\usepackage{multicol}

\usepackage{extarrows}
\usepackage{xcolor}
\usepackage{amsmath, array, amssymb, amsfonts,amsthm}
\usepackage{upgreek}
\usepackage{xfrac}
\usepackage[inline]{enumitem}
\setlist{nolistsep}
\usepackage[french, italian, english]{babel}
\usepackage[all]{xy}
\usepackage{txfonts}  
\usepackage{sectsty}
\usepackage{booktabs}
\usepackage{caption}
\usepackage{dsfont}
\usepackage{mathtools}
\usepackage{slashed}
\usepackage[makeroom]{cancel}
\usepackage[hidelinks]{hyperref}
\usepackage{textcomp}
\usepackage{multicol}
\usepackage[title]{appendix}
\usepackage[square,numbers,sort,compress,merge]{natbib}
\let\cite\citep 
\usepackage{hyperref}
\hypersetup{colorlinks=true, urlcolor=blue, citecolor=blue, linktoc=page}

\usepackage{tikz}
\usepackage{tikz-cd}
\usetikzlibrary{cd}
\tikzcdset{
arrow style=tikz,
diagrams={>={Straight Barb[scale=0.8]}}
}
\usetikzlibrary{matrix,arrows,decorations.pathmorphing}

\DeclareMathAlphabet{\mathpzc}{OT1}{pzc}{m}{it}

\usepackage{fancybox}

\usepackage{mathrsfs}
\usepackage{scrextend}

\makeatletter
\renewcommand*\env@matrix[1][\arraystretch]{%
  \edef\arraystretch{#1}%
  \hskip -\arraycolsep
  \let\@ifnextchar\new@ifnextchar
  \array{*\c@MaxMatrixCols c}}
\makeatother

\newcommand{\defeq}{\vcentcolon=}
\newcommand{\rdefeq}{=\vcentcolon}

\newcommand\M{\mathcal{M}}

\newcommand\RR{\mathbb{R}}
\newcommand\CC{\mathbb{C}}

\newcommand\C{\mathcal{C}}
\renewcommand\1{\textbf{1}}

\renewcommand\H{\mathcal{H}}

\newcommand\E{\mathcal{E}}

\renewcommand\L{\mathcal{L}}
\renewcommand\S{\mathcal{S}}

\newcommand\U{\mathcal{U}}
\newcommand\SO{\mathcal{SO}}
\newcommand\SL{\mathcal{SL}}

\newcommand\K{\mathcal{K}}
\newcommand\J{\mathcal{J}}
\renewcommand\O{\mathcal{O}}
\newcommand\GL{\mathcal{GL}}
\newcommand\D{\mathcal{D}}

\newcommand\n{\text{\tiny{N}}}

\renewcommand\epsilon{\varepsilon}

\newcommand\rarrow{\rightarrow}

\newcommand\LieG{\mathfrak{g}}
\newcommand\LieH{\mathfrak{h}}

\newcommand\so{\mathfrak{so}}

\renewcommand\b{\bar }

\renewcommand\d{\partial}
\newcommand\s{\sigma}
\newcommand\bs{\boldsymbol}

\renewcommand\-{^{-1}}

\newcommand\ad{\text{ad}}

\renewcommand\1{\mathds{1}}

\makeatletter

\newcommand{\Rmnum}[1]{\expandafter\@slowromancap\romannumeral #1@}
\makeatother

\makeatletter
\newcommand{\leqnomode}{\tagsleft@true\let\veqno\@@leqno}
\newcommand{\reqnomode}{\tagsleft@false\let\veqno\@@eqno}
\makeatother

\DeclareMathOperator{\Diff}{Diff}

\DeclareMathOperator{\vol}{vol}

\theoremstyle{definition}

\makeatother

\usepackage{float}

\begin{document}


\title{Relational Supersymmetry {via the Dressing Field Method} and \\ Matter-Interaction Supergeometric Framework}

\author{J. \textsc{François} $\,{}^{a,\,b,\,c,\,*}$ \and L. \textsc{Ravera} $\,{}^{d,\,e,\,f,\,\star}$ }

\date{}

\maketitle
\begin{center}
\vskip -0.6cm
\noindent
{\footnotesize{
${}^a$ Department of Mathematics \& Statistics, Masaryk University -- MUNI.\\
Kotlářská 267/2, Veveří, Brno, Czech Republic.\\[1mm]
 
${}^b$  Department of Philosophy -- University of Graz. \\
 Heinrichstraße 26/5, 8010 Graz, Austria.\\[1mm]
 
${}^c$ Department of Physics, Mons University -- UMONS.\\
20 Place du Parc, 7000 Mons, Belgium.
\\[1mm]

${}^d$ DISAT, Politecnico di Torino -- PoliTo. \\
Corso Duca degli Abruzzi 24, 10129 Torino, Italy. \\[1mm]

${}^e$ Istituto Nazionale di Fisica Nucleare, Section of Torino -- INFN. \\
Via P. Giuria 1, 10125 Torino, Italy. \\[1mm]

${}^f$ \emph{Grupo de Investigación en Física Teórica} -- GIFT. \\
Universidad Cat\'{o}lica De La Sant\'{i}sima Concepci\'{o}n, Concepción, Chile. \\[1mm]


${}^*$ {\small{jordan.francois@uni-graz.at}} \qquad \quad ${}^\star$ {\small{lucrezia.ravera@polito.it}}
}}
\end{center}


\vspace{-3mm}

\begin{abstract}
Relationality is the paradigmatic conceptual core of general-relativistic gauge field theory. It can be made manifest via the \emph{Dressing Field Method} (DFM) of symmetry reduction, a systematic tool to achieve gauge-invariance by extracting the physical degrees of freedom representing relations among field variables. 
We review and further expand on some applications of the DFM to the very foundations of the supersymmetric framework, where it allows to build \emph{relational supersymmetric field theory} and (dis)solves crucial issues.
Furthermore, we elaborate on a novel approach within the relational supersymmetric field theory framework  giving a unified description of fermionic matter fields and bosonic gauge fields, and thus close to Berezin's original motivation for the introduction of supergeometry in fundamental physics: a \emph{Matter-Interaction Supergeometric Unification} (MISU). This new approach stands irrespective from the ultimate empirical status of standard supersymmetric field theory, about which it is agnostic.

\end{abstract}


\vspace{-3mm}

\tableofcontents

\bigskip


\section{Introduction}\label{Introduction}  

A fundamental insight of general-relativistic physics is that diffeomorphisms covariance encodes its  paradigmatic feature: dynamical physical field-theoretical degrees of freedom (d.o.f.) relationally co-define each other, via pointwise
coincidences, in a diffeo-invariant way, thereby defining \emph{physical spacetime} points/events. This is what we
shall call \emph{relationality} hereafter: it implies background independence, i.e. the idea that there are no non-dynamical physical entities, acting upon without being acted upon.
This insight, discovered by articulating  his so-called hole and point-coincidence arguments, was pivotal to Einstein's
completion of General Relativity (GR). See e.g. \cite{Komar1958,Bergmann1961,Norton1987,Norton1988,Stachel1989,Norton1993,Giovanelli2021}, and \cite{JTF-Ravera2024c} for the extension of those arguments to a fully-fledged bundle-geometric framework, in which invariance under the action of the group of bundle automorphisms is discussed and achieved for general-relativistic Gauge Field Theory (gRGFT) for the first time -- both formally and with field-theoretical applications, e.g. to gravity and electromagnetism coupled to scalar fields.  
Relationality, though a core insight of fundamental physics established in the framework of gRGFT, is
seldom recognized as such. Typically, at best, field-theoretical setups exhibit symmetry covariance and are tacitly relational, while better desiderata would be manifest invariance and explicit relationality. The latter, while conceptually very appealing, are more challenging to implement from a technical standpoint unless one has a systematic and mathematically rigorous framework to do so.

In the last ten years, an innovative new approach to reduce gauge symmetry was
developed: the \emph{Dressing Field Method} (DFM), a \emph{systematic tool}\footnote{As far as we know, the only one of the kind, at present.} to build gauge-invariant variables
grounded in the differential geometry of fiber bundles, thus fully non-perturbative. 
{First developed in \cite{GaugeInvCompFields, Francois2014}
in the context of Yang-Mills-type theory and gauge formulations of gravity, i.e. for internal gauge groups, it notably provides an alternative to the Spontaneous Symmetry Breaking interpretation of Higgs-type models, with strong conceptual implications explored in  \cite{Francois2018, Berghofer-et-al2023}, and it modifies the BRST algebra of a gauge theory via the notion of ``dressed ghost", as showcased in \cite{FLM2015_II, FLM2015_I}.
It was also shown to have natural applications in conformal Cartan geometry and twistor theory \cite{ ,Attard-Francois2016_II, Attard-Francois2016_I}, whereby an original mathematical structure was uncovered as a result \cite{Francois2019_II}. These results were reviewed in \cite{Attard_et_al2017}. 
The bundle geometry of the field space of a gauge theory was shown to be the most natural mathematical arena of the DFM in \cite{Francois2021, Francois-et-al2021} -- where the method was also revealed to be the geometric underpinning of the notion of ``edge modes" introduced in the study of the covariant phase space of gauge theories over bounded regions -- while \cite{Zajac2023} further elaborated on its bundle geometric foundations.
The important technical and conceptual distinction between the DFM and gauge-fixing was discussed specifically in \cite{Berghofer-Francois2024} and \cite{Masson-et-al2024} -- the latter reference emphasizing the Faddeev-Popov gauge-fixing of the Path Integral.
The most complete technical and conceptual presentation of the DFM to date is \cite{JTF-Ravera2024gRGFT} where, generalizing \cite{Francois2023-a},  the method is extended to cover the whole framework of classical gRGFT. 
It is in that context that it was shown to supply a natural technical implementation of \emph{relationality}, allowing to reformulate the
foundation of physics in a principled, manifestly relational and automatically invariant way. 
Recently, the method has been shown to apply to (a bundle geometric formulation of) non-relativistic Quantum Mechanics \cite{JTF-Ravera2024NRrelQM}, thus making explicit its relational character, distinct from that meant in Relational QM \cite{Rovelli1996} and exactly analogue to that of gRGFT. 
}

 {Also quite} recently, the DFM has been applied  to the foundations of supersymmetric field theory, in particular to the case of the Rarita-Schwinger (RS) and gravitino fields \cite{JTF-Ravera2024-SUSY}: There it was shown that, 
while usually understood to result from a gauge-fixing, these are actually instances of ``self-dressed", relational variables.\footnote{The standard view is that some ``gauge-fixing" constraint is imposed on them to get the desired number of (off-shell) degrees of freedom. As we will briefly recall below, such constraints, when solved explicitly result not in a gauge-fixing, but in a dressing operation. Calling them ``gauge-fixings" is thus a \emph{wrong statement}, both conceptually and mathematically.   
See {the references mentioned above}, \cite{Berghofer-Francois2024, Masson-et-al2024} and \cite{JTF-Ravera2024gRGFT,JTF-Ravera2024-SUSY}, for details on the clean difference between dressing and gauge-fixing.} 
This fact emerges right from the start, that is from the kinematics one typically considers in supersymmetric field theory, and has deep consequences on the formulation of theories based on supersymmetry (susy), as we will discuss in this work. Moreover, the DFM is at the very root of ``unconventional supersymmetry" (ususy) formulated via the so-called \emph{matter ansatz}, as shown in \cite{JTF-Ravera2024ususyDFM}. 
The understanding of this fact is not only foundational to the three-dimensional ususy model originally proposed in \cite{Alvarez:2011gd} (AVZ model) and to its entire physical and geometric  framework, but it is also the foundation for the formulation of a novel supergeometric setup we here put forth: a \emph{Matter-Interaction Supergeometric Unification} (MISU), in which the framework of supersymmetric field theory is used as a tool to provide a unified description of fermionic matter fields and bosonic gauge fields, as parts of a single superconnection, while remaining agnostic on the ultimate fate of supersymmetry and the existence of particle superpartners.
Our framework is based on  supersymmetrizations of the Lorentz algebra, and the matching between bosonic and fermionic d.o.f. is not required \cite{Sohnius:1985qm}. Furthermore, as we shall discuss, it corrects attempts presented in the literature \cite{Alvarez:2013tga,Alvarez:2021zhh,Alvarez:2023auf,Alvarez:2020qmy} to extend the ususy idea to higher dimensions, providing an unambiguous setup with a solid foundation in differential geometry.

{This contribution is primarily indented as a {focused review paper} of applications of the DFM to foundational aspects of standard supersymmetric {field theory, so as to advertise the approach to a broader community that  hitherto has had little to no exposure to its technical and conceptual underpinning.}
Yet, it also contains the preliminary original results mentioned above concerning the MISU approach.}
It is thus structured as follows: {Sections \ref{Basics of gauge theory and symmetry reduction via the DFM} and \ref{Relational supersymmetry} review fundamental aspects of standard susy through the lens of the DFM.} In Section \ref{Basics of gauge theory and symmetry reduction via the DFM} we provide the basics of gauge field theory and methodically review the key aspects of the DFM, both in the finite case and at the perturbative level. 
We stress the difference between gauge-fixing and the implementation of the DFM via \emph{field dependent dressing fields}. 
We also give the dressed BRST algebra. In Section \ref{Relational supersymmetry} we recall the cases of the RS spinor-vector and gravitino field as self-dressed field-theoretic variables, rather than gauge-fixed objects. 
Then, in Section \ref{Matter-Interaction Supergeometric Framework} we {present new material:} our prescription for a naturally relational Matter-Interaction Supergeometric Framework {and its dressed BRST formulation, which we illustrate via a simple case. 
}
We conclude with Section \ref{Conclusion}, where we comment on these results, highlighting some consequences of our \emph{relational supersymmetric field theory} framework. 
Finally, we discuss the prospects of the novel, intrinsically relational, Matter-Interaction Supergeometric Framework, mentioning also some planned future developments of our findings.
{In Appendix \ref{Perturbatively dressed BRST formalism}, for the sake of {completeness}, we introduce the perturbatively dressed BRST formalism.}

\section{Basics of gauge theory and symmetry reduction via the DFM}\label{Basics of gauge theory and symmetry reduction via the DFM}

In order to better frame and understand the DFM of symmetry reduction, let us start by introducing the typical field content and the associated geometric structure of Gauge Field Theory (GFT), sketching first the kinematics/geometry and then the dynamics.

\paragraph{Kinematics/geometry of GFT} 

A gauge theory is based on a (finite-dimensional) Lie group $H$,\footnote{Typically, considering GFT under a bundle-geometric perspective, $H$ is called the structure group.} with corresponding Lie algebra $\LieH$, and defined over a region $U\subset M$ of a $d$-dimensional manifold $M$.
The basic field content is given by the gauge potential (or connection) 1-form $A={A}_\mu \, dx^{\,\mu} \in \Omega^1(U, \LieH)$ and the matter fields $\phi \in \Omega^\bullet (U,V)$, with $V$ a representation space via $\rho:H \rarrow GL(V)$, and $\rho_*:\LieH \rarrow \mathfrak{gl}(V)$. The minimal coupling prescription is provided by the covariant derivative $D\phi := d\phi + \rho_*(A)\, \phi \in \Omega^{\bullet+1}(U,V)$. The field strength of $A$ (that is the curvature) is $F=dA+\sfrac{1}{2}\,[A,A] \in \Omega^2(U,\LieH)$, and fulfills the Bianchi identity $D^{\,A}F=0$.

The (infinite-dimensional) gauge group of the theory, that is the set of $H$-valued functions $\upgamma: U \rightarrow H$, $x \mapsto \upgamma(x)$, with
point-wise group multiplication $(\upgamma \upgamma')(x)=\upgamma(x) \upgamma'(x)$, defined by
\begin{align}
\label{Gauge-group}
\H := \left\{ \upgamma, \upeta :U \rightarrow H\ |\  \upeta^\upgamma\defeq \upgamma^{-1} \upeta\upgamma\, \right\},
\end{align}
acts on the fields and their field strengths.
The corresponding Lie algebra is 
\begin{align}
\label{LieGaugeGrp}
\text{Lie}\H := \Big\{ \lambda, \lambda' :U \rightarrow \LieH\ |\  \delta_{\lambda}\lambda' \defeq [\lambda', \lambda]\, \Big\}.
\end{align}
The gauge transformations of the fields are defined by the action of $\H$ at the finite level and of Lie$\H$ infinitesimally:\footnote{Observe that in supersymmetric field theory one commonly deals only with the infinitesimal transformations.}
\begin{equation}
\begin{aligned}
\label{GTgauge-fields}
&A\ \mapsto\ A^\upgamma:=\upgamma^{-1} A \upgamma + \upgamma^{-1} d \upgamma , \quad \phi \ \mapsto \ \phi^\upgamma:=\rho(\upgamma)^{-1}\phi,\\[1mm]
&\text{infinitesimally, } \quad \delta_\lambda A = D\lambda=d\lambda +\ad(A) \lambda, \quad \delta_\lambda \phi=-\rho_*(\lambda)\, \phi.
\end{aligned} 
\end{equation}
Consequently, the corresponding field strengths gauge-transform as
\begin{equation}
    \begin{aligned}
        F \ & \mapsto\ F^\upgamma=\upgamma^{-1} F \upgamma , \\
        D\phi \ & \mapsto \ (D\phi)^\upgamma :=\ d\phi^\upgamma + \rho_*(A^\upgamma)\phi^\upgamma =\rho(\upgamma)^{-1}D\phi.
    \end{aligned}
\end{equation}
The action of $\H$ on the \emph{field space} $\Phi=\{A, \phi\}$ of a GFT is a right action, $(A^\upeta)^\upgamma=A^{\upeta\upgamma}$ and $(\phi^\upeta)^\upgamma=\phi^{\upeta\upgamma}$, and foliates $\Phi$ into gauge orbits $\O^\H$. The latter, under adequate restrictions on either $\Phi$ or $\H$, are isomorphic to $\H$ -- cf., e.g., \cite{Singer1978, Singer1981, Ashtekar-Lewandowski1994, Baez1994,  Fuchs-et-al1994, Fuchs1995}. Then, $\Phi$ is a principal fiber bundle with structure group $\H$ over the \emph{moduli space} of orbits $\M\defeq\Phi/\H$. The projection $\Phi \xrightarrow{\pi} \M$, with $\pi(A, \phi)=[A,\phi]$, maps to equivalence classes in the moduli space. 

\paragraph{Gauge fixing}

\begin{figure}[ht]
\begin{center}
\includegraphics[width=0.7\textwidth]{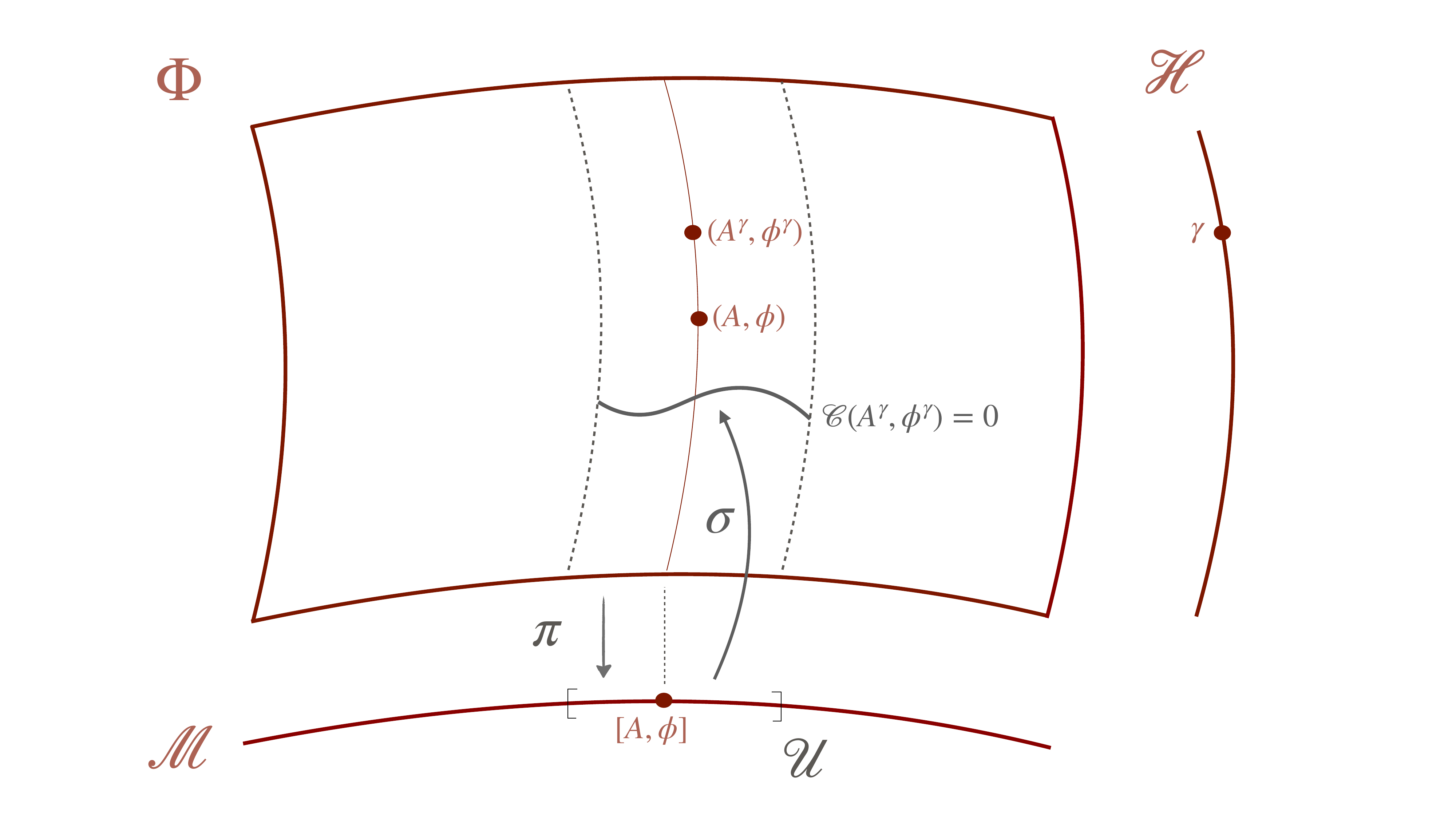}
\caption{Gauge-fixing in $\Phi$, i.e. a choice of local section $\sigma$ of $\Phi$. The gauge-fixing slice is the image of $\sigma$.}
\label{figure:bundlegeomfs}
\end{center}
\end{figure}

It is often considered convenient to somehow restrict to those variables $\{A, \phi\}$ satisfying particular functional properties. This can make computations more manageable and is typically considered a necessary step towards quantization.
When restrictions are imposed by exploiting the gauge freedom \eqref{GTgauge-fields} of the fields, they are referred to as ``\emph{gauge-fixing}" conditions.  
The functional restrictions define a ``slice" in $\Phi$, namely cutting across gauge orbits once. This selects a single representative in each. In other words, as schematically represented in Figure \ref{figure:bundlegeomfs}, a gauge-fixing is a \emph{choice of local section} of the field space bundle $\Phi$, namely $\s: \U\subset\M \rarrow \Phi$.

Concretely, a gauge-fixing is specified by a condition taking the form of an algebraic and/or differential equation on fields, implemented by using the gauge freedom \eqref{GTgauge-fields}: $\C(A^\upgamma, \phi^\upgamma)=0$. 
The gauge-fixing slice is the submanifold
\begin{align}
  \S\defeq \{(A, \phi) \in \Phi_{|\U}\, |\, \C(A^\upgamma, \phi^\upgamma)=0 \} \subset \Phi .
\end{align}
Such a section does not exist globally, meaning that there is no ``good" gauge-fixing, unless $\Phi$ is a trivial bundle, i.e. $\Phi=\M \times \H$ (cf. the literature on the Gribov-Singer obstruction/ambiguity, starting with \cite{Singer1978, Singer1981, Fuchs1995}).
We highlight these features of gauge-fixing to pave the way for a better understanding of the difference between the latter and the dressing via DFM, which will be discussed in the following.

\paragraph{Dynamics of GFT}

The dynamics of a GFT is given by a Lagrangian form
\begin{align}
    L=L(A, \phi) = \L (A,\phi) \vol_d \in \Omega^d(U, \RR),
\end{align} 
with $\vol_d$ the volume form on the region $U\subset M$, with dimension $d$. The Lagrangian is typically required to be $\H$-quasi-invariant (\emph{Gauge Principle}), meaning $\H$-invariant up to boundary terms:
\begin{align}
    L(A^\upgamma, \phi^\upgamma)=L(A, \phi) + db(A, \phi; \upgamma),
\end{align}
for $\upgamma\in \H$, so that the field equations $\bs E(A, \phi)=0$ remain $\H$-covariant. The infinitesimal gauge transformation of the Lagrangian reads
\begin{align}
\label{infgaugetrL}
  \delta_\lambda L(A, \phi) = d\beta(A, \phi; \lambda).
\end{align}

\paragraph{BRST formalism}

In GFT, to the \emph{infinitesimal} generators of gauge transformations -- cf. the 2nd line in \eqref{GTgauge-fields} -- one can associate a Faddeev-Popov
\emph{ghost field} $c$, which is the field-theoretic place holder for the Maurer-Cartan form of the gauge group $\mathcal{H}$. The \emph{BRST algebra} of a (non-Abelian) GFT is defined as (see \cite{Becchi:1975nq})
\begin{equation}
\label{BRSTalgebradef}
\begin{aligned}
    & sA=-Dc \defeq -dc -[A,v], \quad s \phi = -\rho_*(c)\,\phi , \\
    & sc = - \frac{1}{2}[c,c] , 
\end{aligned}
\end{equation}
and we also have, for the curvature $F$ and covariant derivative of the matter field $\phi$,
\begin{align}
\label{curvBRSTalg}
    sF = [F,c] , \quad s(D\phi) = - \rho_*(c)\,D\phi.
\end{align}
The BRST operator $s$ is an antiderivation which anticommutes with
the exterior differential $d$, so that  $sd+ds=0$, and with odd differential forms. 
The bracket $[ \,, \,]$ is  graded  w.r.t. the form and ghost degrees. It is easily verified that $s^2=0$ -- and, since $d^2=0$, we have $(s+d)^2=0$.

Let us also mention that such differential algebra can be recast as a bigraded algebra, with total degree the sum of the form and ghost degrees, whose nilpotent operator is $\tilde{d}\defeq d+s$, such that (s.t.) $\tilde{d}^2=0$.
One define the ``algebraic connection" \cite{Dubois-Violette:1986vtp} $\tilde{A}\defeq A+c$ of bidegree $1$. 
The above BRST algebra follows from what Stora referred to as the ``Russian formula" $\tilde{d}\tilde{A}+\tfrac{1}{2}[\tilde{A},\tilde{A}]=F$, by expanding it w.r.t. the ghost degree. 
The same can be done for the matter sector. In fact, $\phi$ being a 0-form stands alone in the bigraded algebra $\tilde{\phi}=\phi$, and, if
one requires the ``horizontality condition" (see \cite{Baulieu:1985md}) $\tilde{D}\tilde{\phi} \defeq \tilde{d}\tilde{\phi} + \rho_*(\tilde{A})\,\tilde{\phi}=D\phi$, the BRST variation of the matter sector above is recovered.
The BRST transformation of the Lagrangian is
\begin{align}
  s L(A, \phi) = d\beta(A, \phi; c),
\end{align}
which reproduces \eqref{infgaugetrL}. 
A Lagrangian functional thus belong to the $s$ modulo $d$ cohomology: $L \in H^{0,n}(s | d)$, where $0$ is the ghost degree and $n$ the de Rham form degree.

\subsection{The Dressing Field Method}\label{The Dressing Field Method}

The DFM is a systematic tool to produce gauge-invariants out of the field space $\Phi$ of a gauge theory with gauge group $\H$ whose action on $\Phi$ defines gauge transformations, as described above.
Although the DFM is a relatively recent method, it has already been presented multiple times in the literature, either in a more formal and abstract manner or in a more explicit way, depending on the specific context. We refer the interested reader to, e.g., Refs. \cite{Francois2023-a,JTF-Ravera2024gRGFT,JTF-Ravera2024-SUSY,JTF-Ravera2024ususyDFM,JTF-Ravera2024NRrelQM,Berghofer-et-al2023,Berghofer-Francois2024}.
Here, we will review the key aspects of the DFM, which is key to understanding  its applications within the supersymmetric framework to be presented and discussed next.\footnote{The DFM applies indeed gauge theories based on Lie groups and graded Lie groups (supergroups), hence to supersymmetric gauge field theories and supergravity.}
We consider a $\H$-gauge theory with Lagrangian $L(A, \phi) \in \Omega^d(U, \RR)$ and expose how the DFM manifests in its  kinematics first, and then its dynamics. 

\subsubsection*{Kinematics}

Consider a subgroup $K$ of the structure group $H$, $K \subseteq H$, to which corresponds the gauge subgroup $\K \subseteq \H$. 
Suppose there is a group $G$ s.t. either $H\supseteq G \supseteq K$, or $G \supseteq H$. 
A $\K$-\emph{dressing field}  is a map
\begin{align}
    u: M \rarrow G,
\end{align}
i.e. a \emph{$G$-valued field}, defined by its $\K$-gauge transformation: 
\begin{align}
\label{GT-dressing}
u^\kappa:=\kappa\- u, \quad \text{ for } \kappa \in \K.
\end{align} 
The space of $G$-valued $\K$-dressing fields is denoted by $\D r[G, \K]$, where $\K$ (or $K$) is the \emph{equivariance group} of $u$, while $G$ is its \emph{target group}.
Considering the fields $\{A, \phi \}$ as above, given the existence of a $\K$-dressing field we may ``dress" them, thereby defining the  \emph{dressed fields} 
\begin{align}
\label{dressed-fields}
A^u\defeq  u\- A u + u\- du, \quad \phi^u\defeq u\- \phi.
\end{align}  
This illustrates the DFM ``\emph{rule of thumb}": To obtain the dressing of an object, compute first its gauge transformation, then (formally) substitute the gauge parameter with $u$ in the result. Note, however, that the dressing field \emph{is not} an element of the gauge group.
The dressed fields \eqref{dressed-fields} are $\K$-invariant, as is clear from \eqref{GTgauge-fields} and \eqref{GT-dressing}.
Therefore, when $u$ is a $\H$-dressing field, the dressed fields are  $\H$-invariant.
The dressed curvature of $A^u$ is
\begin{align}
    F^u=u\- Fu=dA^u+\sfrac{1}{2}\,[A^u, A^u],
\end{align}
and the dressed covariant derivative is 
\begin{align}
    D^u=d + \rho_*(A^u) , \quad \text{so that} \quad D^u\phi^u =\rho(u)\-D\phi=  d\phi^u + \rho_*(A^u)\phi^u.
\end{align}
The dressed curvature $F^u$ satisfies the Bianchi identity $D^{\,A^u}F^u=0$.

\paragraph{Residual transformations}

Let us remark that, being $\K$-invariant, the dressed fields \eqref{dressed-fields} are expected to display \emph{residual transformations} under what remains of the gauge group. 
If $K$ is a normal subgroup of $H$, $K \triangleleft H$, then $H/K\rdefeq J$ is a Lie group. 
Correspondingly, $\K \triangleleft \H$ and $\J =\H/\K$ is a gauge subgroup of $\H$.
In this case, the dressed fields \eqref{dressed-fields} may exhibit well-defined residual $\J$-gauge transformations, which are called \emph{residual transformations of the 1st kind}. 
If, e.g., the $\K$-dressing field transforms as $u^\upeta =\upeta\- u\, \upeta$ for $\upeta\in \J$,  
the dressed fields are $\J$-gauge variables, satisfying \eqref{GTgauge-fields} with $\upgamma \rarrow \upeta$. 

Dressed objects may also exhibit residual transformations resulting from a possible ``ambiguity" in the choice of dressing field: Two dressing fields $u, u' \in \D r[G, \K]$ may a priori be related by $u'=u\xi$, where $\xi$ is an element of what is referred to as the group of \emph{residual transformations of the 2nd kind}. We may write its action on a dressing field by $u^\xi=u\xi$. 
This group has no action on ``bare" objects. Hence, its action on dressed variables, say dressed forms $\beta^u$, is found via
$(\beta^u)^\xi \defeq (\beta^\xi)^{u^\xi}=\beta^{u\xi}$. Residual transformations of the 2nd kind have been shown to encode \emph{physical reference frame covariance}, both at the classical and at the quantum level \cite{JTF-Ravera2024NRrelQM}. 

\paragraph{Field-dependent dressing fields and DFM relational interpretation}

Key to the DFM is the fact that a dressing field should be extracted \emph{from the field content} of the theory, rather than being introduced in some \emph{ad hoc} way. This means that it has to be a \emph{field-dependent dressing field}, functional on $\Phi$:
\begin{equation}
\label{Field-dep-dressing}
\begin{aligned}
u\ \ :\ \  \Phi \ &\rarrow\  \D r[G, \K], \\
    \{A, \phi\} \  &\mapsto\  u=u[A, \phi] .
\end{aligned} 
\end{equation}
Note that the original fields $\{A, \phi\}$ encode redundant d.o.f., in the sense that physical d.o.f. are mixed with non-physical pure gauge modes. We may understand the dressed fields $\{A^{u[A, \phi]}, \phi^{u[A, \phi]}\}$ as a reshuffling of the d.o.f. of the original fields. This implies an elimination, partial or complete, of the pure gauge modes. 

In the presence of field-dependent dressing fields \eqref{Field-dep-dressing}, the DFM has a natural \emph{relational} interpretation \cite{JTF-Ravera2024c,Francois2023-a}:
The dressed fields $\{A^{u[A, \phi]}, \phi^{u[A, \phi]}\}$ represent the gauge-invariant, physical \emph{relations} among  d.o.f. embedded in the original (bare) fields $\{A, \phi\}$.\footnote{Observe that this can also occur among the d.o.f. of $A$ and $\phi$ themselves, in which case we talk about ``self-dressings", $A^{u[A]}$ or $\phi^{u[\phi]}$.}

\paragraph{Perturbative dressing}

It may be -- as it is typically the case in supersymmetric field theory -- that one is interested in \emph{invariance at first order}, that is under the \emph{infinitesimal} gauge transformations in  \eqref{GTgauge-fields}.
Let us keep  in mind that geometrically this means transformations that are not only linear in the gauge parameter  $\lambda$ (i.e. 1st order \emph{in perturbation theory}), but well-defined in the relevant mathematical space. 
In this case, in the implementation of the DFM one is led to define an infinitesimal Lie$\K$-dressing field  as
\begin{align}
\label{pert-dressing-field}
\upsilon: U\subset M \rarrow \LieG, 
 \quad \text{s.t. } \quad \delta_\lambda \upsilon \approx -\lambda \ \ \text{for}\ \  \lambda \in \text{Lie}\K,
\end{align}
neglecting higher-order terms,  polynomial in $\lambda$ and $\upsilon$ yet not in $\LieG$, in the first order transformation property.\footnote{Notice that residual $\J$-transformations of the 1st kind of the form $u^\upeta=\upeta^{-1} u \,\upeta$ with $\upeta \in \J$ have linear version  $\delta_\theta \upsilon = [\upsilon,\theta]$ with $\theta \in$ Lie$\J$. This relation may seem to be of order 2 in perturbation theory, since both $\theta$ and $\lambda$ are linear parameters, but it is geometrically of first order as both sides are $\LieG$-valued.}
 The~\emph{perturbatively dressed fields} are then defined as,
\begin{align}
\label{pertdrfieldsgendef}
    \upphi^\upsilon\defeq \upphi + \b\updelta_\upsilon \upphi, \quad \upphi=(A, \phi),
\end{align}
where $\b\updelta_\upsilon \upphi$  ``mimics" the functional expression of the infinitesimal gauge  transformation $\delta_\lambda \upphi$ in \eqref{GTgauge-fields}, with $\lambda \rarrow \upsilon$ (so, in no sense is $\b\updelta_\upsilon$ a differential of the  algebra of fields).
More explicitly, we have the perturbatively dressed gauge potential and matter field
\begin{equation}
\label{pert-dressed-fields}
\begin{aligned}
A^\upsilon\defeq&\ A+ \b\updelta_\upsilon A   \quad
&&\phi^\upsilon \defeq \phi+ \b\updelta_\upsilon \phi
\\
=& \, A+ D\upsilon,   &&\phantom{\phi...}= \phi -\rho_*(\upsilon)\,\phi.
\end{aligned}
\end{equation}
These  perturbatively dressed fields are $\K$-invariant \emph{at 1st order}:
\begin{align}
\delta_\lambda \upphi^\upsilon 
= \delta_\lambda \upphi + \b\updelta_{\delta_\lambda \upsilon} \upphi 
= \delta_\lambda \upphi + \b\updelta_{-\lambda} \upphi
= \delta_\lambda \upphi - \delta_\lambda \upphi \equiv 0,
\end{align}
where one must neglect higher-order terms in $\lambda$ and $\upsilon$, starting from $\b\updelta_\upsilon (\delta_\lambda \upphi)$.
For example, for the perturbatively dressed  gauge potential and the matter field we have
\begin{align*}
    & \delta_\lambda A^\upsilon 
    = \delta_\lambda  A + \delta_\lambda D\upsilon 
    = D\lambda +D (\delta_\lambda \upsilon) + [\delta_\lambda A, \upsilon] 
    =D\lambda + D(-\lambda) 
    + \cancelto{\,\text{\tiny{neglect}}}{[D\lambda, \upsilon]}\!\!
    \equiv 0 , \\
    & \delta_\lambda \phi^\upsilon = 
    \delta_\lambda  \phi - \delta_\lambda \big( \,\rho_*(\upsilon)\, \phi \, \big)
    = -\rho_*(\lambda)\, \phi - \rho_*(\delta_\lambda \upsilon)\, \phi - \rho_*(\upsilon)\,\delta_\lambda \phi
    =-\rho_*(\lambda)\, \phi - \rho_*(-\lambda)\, \phi - \cancelto{\,\text{\tiny{neglect}}}{ \rho_*(\upsilon)\rho_*( -\lambda)\, \phi}\!\!\!
    \equiv 0.
\end{align*}
Let us now address the impact of the DFM on the dynamics of a gauge theory.

\subsubsection*{Dynamics}

Let us consider the quasi-invariant Lagrangian form $L=L(A, \phi)$ of an $\H$-gauge theory.
Supposing that there exists a $\K$-dressing field $u$ with target group $G\subseteq H$, 
we may  exploit the quasi-invariance of $L$ to define the \emph{dressed Lagrangian} as the Lagrangian expressed in terms of the dressed fields in \eqref{dressed-fields}:
\begin{equation}
\label{dressed-Lagrangian}
L(A^u, \phi^u)=L(A, \phi) + db(A, \phi; u).
\end{equation}
It is  easily obtained from the quasi-invariance of Lagrangian and using the DFM rule of thumb.
Notice that if $L$ is strictly $\H$-invariant, i.e. s.t. $b=0$, then $L(A, \phi)= L(A^u, \phi^u)$. 
In either cases, the field equations $\bs E(A^u, \phi^u)=0$ for the dressed fields have the \emph{same functional expression} as the field equations for the bare fields, $\bs E(A, \phi)=0$.
Let us also remark that, in the presence of residual $\J$-transformations of the 1st kind, $L(A^u, \phi^u)$ is a $\J$-theory. 

Considering the perturbative dressing scenario described above, for a quasi-invariant Lagrangian, such that 
$\delta_\lambda L(A, \phi)= d\beta(A, \phi; \lambda)$, we may define the perturbatively dressed Lagrangian as
\begin{align}
\label{pert-dressed-Lagrangian}
  L(A^\upsilon, \phi^\upsilon)\defeq L(A, \phi) + d\beta(A, \phi; \upsilon). 
\end{align}
The dressed field equations $\bs E(A^\upsilon, \phi^\upsilon)=0$ in this case are thus $\K$-invariant at 1st order.
Naturally, \eqref{pert-dressed-fields}-\eqref{pert-dressed-Lagrangian} may be obtained by linearization of \eqref{dressed-fields}-\eqref{dressed-Lagrangian}.

\subsubsection*{Difference between dressing via the DFM and gauge-fixing}

Comparing the definition \eqref{Gauge-group} of the gauge group and that of a dressing field \eqref{GT-dressing}, is evident that $u \notin \K$. 
{ 
It follows that a $\upphi$-dependent dressing field $u=u[\upphi]$, which by definition transforms as $u[\upphi]^\kappa:=u[\upphi^\kappa]=\kappa\- u[\upphi]$, 
cannot be misconstrued as a $\upphi$-dependent gauge group element $\gamma[\upphi]$, which by definition is s.t. $\gamma[\upphi]^\kappa:=\gamma[\upphi^\kappa]=\kappa\- \gamma[\upphi] \kappa$.}
This is a crucial fact of the DFM: Despite the formal analogy with \eqref{GTgauge-fields}, the dressed fields \eqref{dressed-fields} are not gauge transformations.
The dressed fields $\{A^u, \phi^u\}$ \emph{are not} a point in the gauge $\K$-orbit $\O^\K_{\{A, \phi\}} \subset \O^\H_{\{A, \phi\}}$ of $\{A, \phi\}$. Hence, $\{A^u, \phi^u\}$ \emph{must not} be confused with a gauge-fixing of the bare variables $\{A, \phi\}$, namely with a point on a gauge-fixing slice $\S$. 
Contrary to a gauge-fixing $\Phi \rarrow \S \subset \Phi$, the ``dressing operation"  is not a map from $\Phi$ to itself, but from $\Phi$ to the space of dressed fields $\Phi^u$, only isomorphic to a subbundle of $\Phi$. 

\begin{figure}[ht]
\begin{center}
\includegraphics[width=0.7\textwidth]{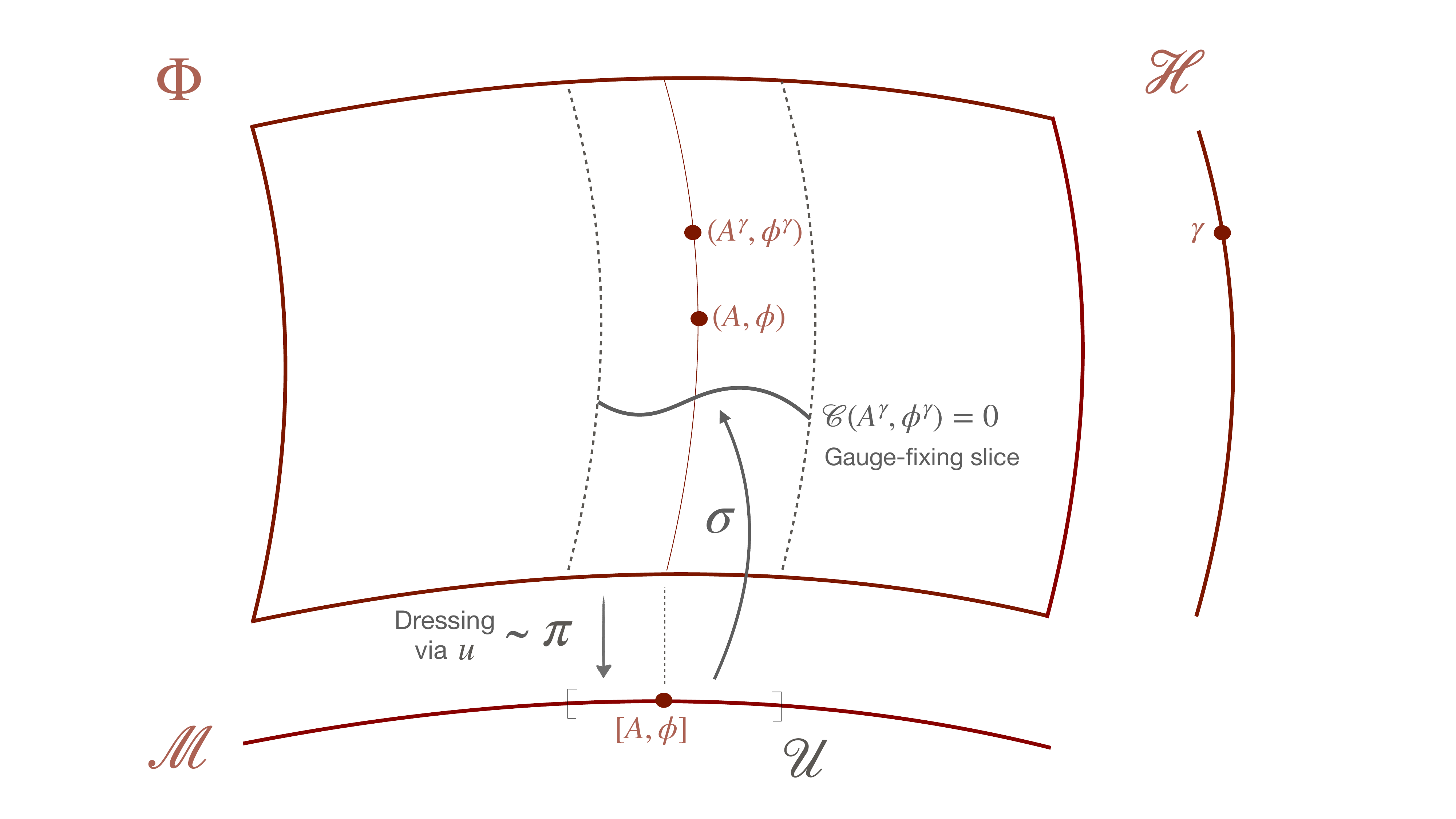}
\caption{Difference between dressing as done via the DFM and gauge-fixing in field space $\Phi$.}
\label{figure:DFMvsgaugefixing}
\end{center}
\end{figure}

In particular, if one considers a complete symmetry reduction via an $\H$-dressing field, then $\Phi^u$ is readily understood as a \emph{coordinatization of the moduli space} $\M$ -- or of a region $\U \subset \M$ over which the field-dependent dressing field $u: \Phi_{|\U} \rarrow \D r[G, \H]$ is defined: the one-to-one mapping $(\Phi_{|\U})^u \leftrightarrow \U \subset \M$ may indeed be seen as a ``coordinate chart". 
It follows that performing the dressing procedure allows to work with the physical d.o.f. which are not accessible in any way to direct computations through gauge-fixing. 
Furthermore, as highlighted above, dressed fields $\upphi^u=\{A^u, \phi^u\}$ are \emph{invariant relational variables}, while representatives fields located on a gauge-fixing slice $\S$ are neither. 
This distinction applies obviously to gauge-fixings whichever ways they are implemented; be it in the most straightforward way, by hand, or via more subtle means like the so-called BRST gauge-fixing (and Faddeev-Popov trick).\footnote{Sometimes called ``covariant" BRST gauge-fixing. Yet the adjective ``covariant" does no relevant work for the purpose of our discussion, as it merely means relativistic (most often even just Lorentz) covariance, which is taken for granted in serious geometric field theory.}
The difference between the implementation of a (full) dressing operation via the DFM and a gauge-fixing in the field space bundle $\Phi$ is schematically represented in Figure \ref{figure:DFMvsgaugefixing}. 

Finally, we remark that the theory
confronted to experimental tests is always the dressed one, not a gauge-fixed version of it, as is often said -- cf. \cite{JTF-Ravera2024gRGFT} for details on this crucial point, and to \cite{Berghofer-Francois2024,Masson-et-al2024} for further details discussion of the technical and conceptual distinction between gauge-fixing and dressing.

\subsection{Dressed BRST formalism}\label{Dressed BRST algebra}

We conclude this section by reviewing the \emph{dressed BRST formalism} in GFT, applications of which to conformal Cartan geometry and twistor theory may be found in  \cite{FLM2015_II,FLM2016_I}.

Given the definition of the ``bare" BRST algebra \eqref{BRSTalgebradef}-\eqref{curvBRSTalg} and of the dressed fields \eqref{dressed-fields}, it is easy to show that the latter satisfy a dressed BRST algebra
\begin{equation}
\label{dressed-BRSTalgebradef}
\begin{aligned}
    & sA^u=-D^u c^u , \quad s \phi^u = -\rho_*(c^u)\,\phi^u, \quad 
     sF^u = [F^u,c^u] , \quad s(D^u\phi^u) = - \rho_*(c^u)\,D^u\phi^u, \\
    & sc^u = - \frac{1}{2}[c^u,c^u].
\end{aligned}
\end{equation}
with the \emph{dressed ghost} $c^u$ defined as
\begin{align}
    \label{dressed-ghost}
    c^u \defeq u\- c u + u\- s u .
\end{align}
These results are  formal, leaving the BRST variation $su$ of the field $u$ unspecified, i.e. independent of wether or not it is a dressing field. 

If it is a $\H$-dressing field indeed, then it satisfies the BRST version of the defining property of a dressing field:  $su=-cu$. 
So the dressed ghost is 
\begin{align}
    c^u \defeq u\- c u + u\- s u = u\- c u + u\- (-cu) \equiv 0,
\end{align}
which trivializes the dressed BRST algebra:
\begin{align}
    sA^u = 0 , \quad s \phi^u =0 , \quad sF^u=0 , \quad s(D^u \phi^u) =0.
\end{align}
thereby expressing that the $\H$ symmetry is totally reduced.
In other words, invariant, dressed variables $\upphi^u$ are in the kernel of the differential BRST operator $s$. 
Therefore, the dressed Lagrangian is $s$-closed: $sL(A^u,\phi^u)=0$.

If $u$ is a $\K$-dressing field,  as we have previously discussed, one expects   residual $\J$-gauge symmetry of the 1st kind. Following the notation introduced previously, we may write
\begin{equation}
\label{splittingresghostands}
\begin{aligned}
    & c = c_K + c_J , \\
    & s = s_K + s_J.
\end{aligned}
\end{equation}
From the defining property $s_K u=-c_K u$ f the dressing field follows that the dressed ghost is
\begin{equation}
\label{dressed-ghost-res1stkind}
\begin{aligned}
    c^u & = u\- (c_k+c_J) u + u\- s u \\
    & = u\- c_K u + u\- c_J u + u\-(-c_K u) + u\- s_J u \\
    & = u\- c_J u + u\- s_J u,
\end{aligned}
\end{equation}
so that the dressed BRST algebra encodes 
 the residual $\J$-symmetry of the $\K$-invariant dressed fields $\upphi^u$.
The explicit form of $s_J u$ depends on the specific case under analysis.
If, for example, the $\K$-dressing field transforms as $u^\upeta =\upeta\- u\, \upeta$ for $\upeta\in \J$, so that the corresponding BRST variation is $s_J u = [u,c_J]$, we see that the dressed ghost \eqref{dressed-ghost-res1stkind} simply reduces to $c^u = c_J$. So that  \eqref{dressed-BRSTalgebradef}  becomes
\begin{equation}
\begin{aligned}
    & sA^u=-D^u c_J , \quad s \phi^u = -\rho_*(c_J)\,\phi^u , \\
    & sF^u = [F^u,c_J] , \quad s(D^u\phi^u) = - \rho_*(c_J)\,D^u\phi^u, \\
    & sc_J = - \frac{1}{2}[c_J,c_J] ,
\end{aligned}
\end{equation}
showing that the  $\K$-invariant dressed fields are standard $\J$-gauge fields: on them the BRST operator reduces to $s=s_J$. Then the BRST variation of the dressed Lagrangian is $sL(A^u,\phi^u)=s_J L(A^u,\phi^u)=d\beta(A^u,\phi^u;c_J)$, i.e. it belongs to the $s_J$ modulo $d$ cohomology: $L^u \in H^{0|n}(s_J|d)$.

In Appendix \ref{Perturbatively dressed BRST formalism} we give the perturbative version of the above, i.e. the \emph{perturbatively dressed BRST algebra} satisfied by perturbatively dressed fields \eqref{pertdrfieldsgendef}-\eqref{perturb-dressed-BRSTalgebradef}.

\section{Relational supersymmetry}\label{Relational supersymmetry}

As shown  first  in \cite{JTF-Ravera2024-SUSY}, two of the most common ``gauge-fixing choices" used in supersymmetric field theory to achieve the desired number of off-shell d.o.f. are actually instances of the DFM. They can be applied to both the Rarita-Schwinger (RS) spinor-vector and the gravitino field, which shows them to be self-dressed, relational, field variables: hence the terminology \emph{relational supersymmetry} we introduce.\footnote{We observe that a \emph{fully} relational theory would have to have all its symmetries reduced via the DFM. In order to do so, one has to first reduce all ``internal" (gauge) symmetries and finally (super)diffeomorphisms.} 
Via the DFM, supersymmetry is therefore reduced, yielding susy-invariant objects (field variables that are singlets under susy). 
The procedure applies at the kinematical level, dynamical considerations coming only after. 
It holds the same in any dimensions and in the presence of any amount $\mathcal{N}$ of susy charges. For simplicity, in the following we will showcase the simple $\mathcal{N}=1$, four-dimensional case. 

\subsection{Rarita-Schwinger spinor-vector self-dressing}\label{Rarita-Schwinger spinor-vector self-dressing}

The Rarita-Schwinger (RS) field in supersymmetric field theories is considered to be a spinor-vector ${\psi^\alpha}_{\!\mu}$,\footnote{Let us mention that the fields introduced by Rarita and Schwinger in the original work \cite{Rarita:1941mf} are more general spinor-tensors of the kind ${\psi^\alpha}_{\,\mu_1\ldots \mu_l}={\psi^\alpha}_{\,(\mu_1\ldots \mu_l)}$, commonly said to describe spin-$(l + \sfrac{1}{2})$ fields on $U \subset \RR^{4}$. They fulfill a (massive) Dirac equation $(\slashed{\d}+m){\psi^\alpha}_{\,\mu_1\ldots \mu_l}=0$, where $\slashed{\d}:=\gamma^{\,\mu}\d_{\mu}$, with $\gamma^{\,\mu}$ the Dirac gamma-matrices, together with the Lorentz-irreducibility condition $\gamma^{\,\mu_1}{\psi^\alpha}_{\,\mu_1\ldots \mu_l}=0$ -- i.e. the gamma-tracelessness condition. 
Note that both the Dirac and the gamma-tracelessness conditions are \emph{on-shell constraints} (i.e., field equations) of the family of Lagrangians introduced by Rarita and Schwinger.
The case $l=1$ correspond to the RS spinor-vector field ${\psi^\alpha}_{\,\mu}$, typically said to carry spin $\sfrac{3}{2}$. 
This is the field we are interested in, which is in fact the one usually considered in supersymmetric field theory.} component of a spinor-valued 1-form field $\psi = \psi_\mu \, dx^{\,\mu} \in \Omega^1(U,\sf S)$, with $U \subset M$ a 4-dimensional manifold and $\sf S$ a (Dirac) spinor representation for the Lorentz group $S\!O(1,3)$. For notational convenience, we will frequently omit the spinor index $\alpha$.
We assume the Majorana condition $\bar \psi = \psi^\dagger \gamma^0 = \psi^t C$, with $C$ the charge conjugation matrix (s.t. $C^t=-C$).
In the context of supersymmetric field theory, the Lagrangian referred to as the RS term is the (massless) theory
\begin{align}
\label{RSLagr}
    L_{\text{RS}}(\psi) = \bar \psi \wedge \gamma_5 \gamma \wedge d \psi \quad \rarrow \quad
    \L_{\text{RS}}(\psi) = \epsilon^{\,\mu \nu \rho \sigma} \bar \psi_\mu \gamma_5 \gamma_\nu \d_\rho \psi_\sigma \,,
\end{align}
with $\gamma:=\gamma_\mu \, d x^{\,\mu}$ the gamma-matrix 1-form. The field equations are\footnote{We remark that this  theory does not coincide with that originally introduced by Rarita and Schwinger in \cite{Rarita:1941mf}. The massless limit of their field equations does not yield \eqref{RSfieldeqs} and the variation of \eqref{RSLagr} does not imply the gamma-tracelessness condition.}
\begin{align}
\label{RSfieldeqs}
    \gamma_5 \gamma \wedge d \psi = 0 \quad \rarrow \quad
    \epsilon^{\,\mu \nu \rho \sigma} \gamma_5 \gamma_\nu \d_\rho \psi_\sigma = 0 .
\end{align}
The Lagrangian \eqref{RSLagr} is quasi-invariant under the susy gauge transformation
\begin{equation}
\begin{aligned}
\label{susygaugetrRS}
    & \psi \mapsto \psi^\upepsilon=\psi + d \upepsilon \quad \rarrow \quad \psi_\mu \mapsto \psi^\upepsilon_\mu=\psi_\mu + \d_\mu \upepsilon , \\
    & \text{infinitesimally, } \quad \delta_\epsilon \psi = d \epsilon \quad \rarrow \quad \delta_\epsilon \psi_\mu = \partial_\mu \epsilon ,
\end{aligned}
\end{equation}
where  $\epsilon$ is the linearisation of
the spin-$\sfrac{1}{2}$ Majorana spinor $\upepsilon=\upepsilon(x)$ belonging to the Abelian (additive) gauge group 
\begin{align}
\label{stranslation-gauge-grp}
\E\defeq\Big\{\upepsilon, \upepsilon'\!:\!U \rarrow T^{0|4}\, |\, \upepsilon^{\upepsilon'}=\upepsilon\, \Big\}, 
\end{align}
where $T^{0|4}\subset T^{4|4}$ is the \emph{supertranslation} subgroup  of the ($\mathcal{N}=1$) super-Poincaré group $sIS\!O(1,3) \defeq S\!O(1,3) \ltimes T^{4|4}$ \cite{Gursey1987,DeAzc-Izq}.
Note that we have
\begin{align}
\left(\psi^\upepsilon\right)^{\upepsilon'}= \left( \psi + d \upepsilon \right)^{\upepsilon'} = \psi^{\upepsilon'} + d\upepsilon^{\upepsilon'} = \psi + d\upepsilon' + d \upepsilon = \psi + d (\upepsilon + \upepsilon') = \psi^{\upepsilon + \upepsilon'}.
\end{align}
We have then the susy transformation of the the Lagrangian
\begin{align}
L_\text{RS}(\psi^\upepsilon) 
= L_\text{RS}(\psi) + db(\psi; \upepsilon)
=L_\text{RS}(\psi) + d(\bar \upepsilon \wedge \gamma_5 \gamma \wedge d \psi ).
\end{align}

In the susy literature, the field   field ${\psi^\alpha}_{\!\mu}$ is usually said to contain a spin-$\sfrac{3}{2}$ and a spin-$\sfrac{1}{2}$ part. 
Yet, in terms of \emph{irreducible} spin representations, it actually involves  a spin-$\sfrac{3}{2}$ component and two spin-$\sfrac{1}{2}$ parts  \cite{VanNieuwenhuizen:1981ae}: 
\begin{align}
    (1\oplus 0)\otimes \sfrac{1}{2}=\sfrac{3}{2}\oplus \sfrac{1}{2}\oplus \sfrac{1}{2}.
\end{align}
The spin-$\sfrac{1}{2}$ fields in this irreducible decomposition corresponds to the gamma-trace and the divergence of ${\psi^\alpha}_{\!\mu}$. 
The decomposition most commonly mentioned in the susy literature,
\begin{align}\label{gammatracedec}
    {\psi^\alpha}_{\!\mu} (\uprho,\chi) := {\uprho^\alpha}_{\!\mu} + \gamma_\mu \, \chi^\alpha ,
\end{align}
where $\chi^\alpha := 1/d \, \gamma^{\,\mu} \, {\psi^\alpha}_{\!\mu}$ is a spin-$\sfrac{1}{2}$ field, is \emph{reducible}. Here ${\uprho^\alpha}_{\!\mu}$ is s.t. $\gamma^{\,\mu}\uprho_\mu =0$, and it contains both a ``longitudinal" mode (divergence-free) and a ``transverse'' mode, 
${\uprho^{\alpha|\text{L}}}_{\!\mu}$ and 
${\uprho^{\alpha|\text{T}}}_{\!\mu}$; they correspond, respectively, to 8 and 4 off-shell d.o.f., namely to a 
spin-$\sfrac{3}{2}$ contribution and a spin-$\sfrac{1}{2}$ part. 
It is commonly said that the spin-$\sfrac{1}{2}$ part of $\psi$ can be eliminated by ``gauge-fixing", typically
\begin{align}
\label{gammatr}
    \gamma^{\,\mu} \psi_{\mu} = 0 ,
\end{align}
and that one is thus left with a spin-$\sfrac{3}{2}$ field (we have $16-4=12$ \emph{off-shell} d.o.f.). 
Actually, these $12$ off-shell d.o.f. correspond to those carried by the residual spin-$\sfrac{3}{2}$ and spin-$\sfrac{1}{2}$ parts after ``gauge-fixing". 
But what is even more important is the fact that, as we will show below, the condition \eqref{gammatr} can be invariantly implemented through symmetry reduction via dressing.  
Therefore, what is commonly referred to as the RS field $\psi$ is in fact obtained via the DFM as a \emph{self-dressed susy-invariant} variable.

Before showing this, let us just recall that taking together the field equations \eqref{RSfieldeqs} and the condition \eqref{gammatr}, one also obtains the transversality condition
\begin{align}
\label{transvcond}
    \d^{\,\mu} \psi_\mu = 0 ,
\end{align}
which therefore follows only \emph{on-shell} in the theory at hand. 
Therefore, $\psi$ satisfies $\Box \psi = 0$, where $\Box:=\d^{\,\mu} \d_\mu$. It describes, \emph{on-shell}, a spin-$\sfrac{3}{2}$ massless particle propagating in a Minkowski background.
Indeed, in the \emph{flat} case, using $[\d_\mu,\gamma_\mu]=0$ together with the properties of the gamma-matrices, it can be shown that the field equations \eqref{RSfieldeqs} imply $\slashed\d (\gamma^{\,\mu} \psi_\mu)-\d^{\,\mu}\psi_\mu=0$, which, taking into account \eqref{gammatr} -- choice made at the kinematical level -- yields \eqref{transvcond}. 
Using the same field equations \eqref{RSfieldeqs} together with the alternative choice \eqref{transvcond}, one  gets the weaker constraint $\slashed\d (\gamma^{\,\mu}\psi_\mu)=0$.

\paragraph{Gamma-trace dressing}

The ``gamma-trace gauge-fixing" can actually be seen as an instance of the DFM. Let us consider the RS field ${\psi^\alpha}_{\!\mu}$  carrying 16 off-shell degrees of freedom, and let us proceed point by point in our analysis: 
\begin{enumerate}
    \item The susy transformation  \eqref{susygaugetrRS} of the reducible gamma-trace decomposition \eqref{gammatracedec}, in dimension $d$, is
    \begin{equation}
    \begin{aligned}
    \label{chirhogaugetr}
    \chi & \,\mapsto\, \chi^\upepsilon= \chi + \tfrac{1}{d} \slashed\d \upepsilon , \\
    \uprho_\mu & \,\mapsto \,\uprho^\upepsilon_\mu = \uprho_\mu - \tfrac{1}{d} \gamma_\mu \slashed \d \upepsilon + \d_\mu \upepsilon .
    \end{aligned}
    \end{equation}
    \item We pick the gamma-tracelessness constraint \eqref{gammatr} as a functional condition on the variable
    \begin{align}
        \psi_\mu^u\defeq \psi_\mu+ \d_\mu u
    \end{align}
    and solve it explicitly for the ``parameter" $u$:
    \begin{equation}
    \label{gamma-tr-dressing}
    \gamma^{\,\mu} \psi^u_\mu = \gamma^{\,\mu} (\psi_\mu + \d_\mu u) = 0 \quad \Rightarrow \quad u[\psi] = - \slashed{\d}\- (\gamma^{\,\mu} \psi_\mu) = - d \slashed{\d}\- \chi.
    \end{equation}
    \item We now have to assess if $u$ is 
    \begin{itemize}
        \item[A)] an element of the gauge group, and thus the gamma-trace constraint \eqref{gammatr} a gauge-fixing;
        \item[B)] a dressing field, in which case $u$ has to transform accordingly.
    \end{itemize}
    For \eqref{gammatr} to be a gauge-fixing, $u$ must be an element of the gauge group $\E$: i.e. it must be gauge-invariant: $u[\psi]^\upepsilon:=u[\psi^\upepsilon]=u[\psi]$ -- because the gauge group $\E$ of supertranslation is Abelian \eqref{stranslation-gauge-grp}. But we can easily check that, as a functional of $\psi$, under the susy transformations \eqref{chirhogaugetr} $u$ gauge transforms as
    \begin{align}
    \label{udressfieldtrepsRS1}
    u[\psi]^\upepsilon \defeq u[\psi^\upepsilon] 
    =  - d \slashed{\d}\- \chi^\upepsilon 
    =- d \slashed{\d}\- \left(\, \chi + \tfrac{1}{d} \slashed\d \upepsilon \right) 
    =- d \slashed{\d}\- \chi - \upepsilon
    =u[\psi] - \upepsilon.
    \end{align}
    The latter is, in fact, the Abelian (additive) version of a \emph{dressing field} transformation \eqref{GT-dressing}. 
    We thus conclude that the explicit solution of the gamma-trace constraint \eqref{gammatr} does not result in a gauge-fixing but in a dressing operation, in the precise technical sense stemming from the DFM. Observe that this dressing is \emph{non-local}. 
    \item The variable $\psi^u$ is thus a  susy-invariant dressed field, expressed in terms of bare fields as:
    \begin{align}
    \label{dressed-RS}
    \psi_\mu^u\defeq&\, \psi_\mu+ \d_\mu u [\psi]
    = \psi_\mu   - d \d_\mu  \slashed{\d}\- \chi .
    \end{align}
    It satisfies $\gamma^{\,\mu}\psi^u_\mu \equiv 0$ by construction. In other words, once $u=u[\psi]$ has been properly extracted as a dressing field, then $\gamma^{\,\mu} \psi^u_\mu \equiv 0$ is trivially satisfied by $\psi^u_\mu$, meaning that this equation is not a constraint for the dressed field $\psi^u_\mu$, just an identity it fulfills. 
    \item We may finally have a look at the decomposition \eqref{gammatracedec} applied to $\psi^u_\mu$. We find the dressed d.o.f.
    \begin{equation}
    \begin{aligned}
    \label{chirhogammatrdressed}
    \chi^u & = \chi + \tfrac{1}{d} \slashed\d u[\psi] = \chi + \tfrac{1}{d} (-d) \slashed{\d} \slashed{\d}\- \chi \equiv 0 , \\
    \uprho^u_\mu & = \uprho_\mu - \tfrac{1}{d} \gamma_\mu \slashed \d u[\psi] + \d_\mu u[\psi] = \psi_\mu^u.
    \end{aligned}
    \end{equation}
\end{enumerate}
In conclusion, in this case the dressing field is given in terms of the gamma-trace component $\chi$, which carries 4 off-shell d.o.f.; the resulting \emph{dressed field} $\psi^u = \uprho^u$ carries 12 physical off-shell d.o.f. ($16-4=12$, as $\chi^u=0$). Note that it still contains both a spin-$\sfrac{3}{2}$ (8 d.o.f.) part and a spin-$\sfrac{1}{2}$ (4 d.o.f.) component.
It should be stressed that in the DFM treatment, the 12 off-shell d.o.f. of the dressed RS field \eqref{dressed-RS} are obtained in a \emph{susy-invariant} way, without any restriction on the gauge group.\footnote{The dressing operation we have just performed is  analogous to the  ``\emph{axial gauge}" for a vector gauge potential $A_{\mu}$ in QED: $n^{\,\mu}A_{\mu}=0$, 
with $n^{\,\mu}$ a constant 4-vector -- cf. e.g. \cite{Leibbrandt-Richardson1992}. 
In fact, as above, by solving explicitly the axial constraint one ends-up building  a (non-local) dressing field $u[A, n]$; the latter then yields the $\U(1)$-invariant (self-)dressed field $A^u_\mu\defeq A_\mu+ \d_\mu u[A, n]$.} 
On the other hand, if one aims to restrict to this 12 d.o.f. by imposing \eqref{gammatr} \emph{by fiat}, this requires a restriction of the field space to the subspace $\psi_\mu \rarrow \uprho_\mu$, and correspondingly a restriction of the gauge group to those elements satisfying $\slashed{\d}\upepsilon=0$ -- by the 1st line of \eqref{chirhogaugetr}.

\paragraph{Transverse dressing}

We repeat the above template, this time considering the alternative functional constraint \eqref{transvcond}, which is (again) usually understood as a gauge-fixing. 
We show that, when solved explicitly, it again yields a dressing field as technically defined in the DFM. 
\begin{enumerate}
    \item We perform the reducible (non-local) decomposition
    \begin{align}
    \label{TLdivergdec}
    \psi_\mu = \psi_\mu^{\text{T}} + \psi_\mu^{\text{L}} \defeq \psi_\mu^{\text{T}} + \d_\mu [ \Box\- (\d^{\,\nu} \psi_\nu) ]  = \psi_\mu^{\text{T}} + \partial_\mu \kappa ,
    \end{align}
    where $\psi^\text{T}_\mu$ is the transverse component and $\psi^\text{L}_\mu:=\d_\mu \kappa$ is the longitudinal $d$-exact component; $\kappa$ is a spin-$\sfrac{1}{2}$ field carrying 4 d.o.f. off-shell.
    Under the susy transformation \eqref{susygaugetrRS} we have
    \begin{equation}
    \begin{aligned}
    \label{psiTkappagaugetr}
    \psi_\mu^{\text{T}} & \,\mapsto\, \big(\psi_\mu^{\text{T}}\big)^\upepsilon= \psi_\mu^{\text{T}} , \\
    \psi_\mu^{\text{L}} & \,\mapsto \,\big(\psi_\mu^{\text{L}}\big)^\upepsilon= \psi_\mu^{\text{L}} + \d_\mu \upepsilon , \quad \text{that is} \quad \kappa \,\mapsto \,\kappa^\upepsilon = \kappa + \upepsilon .
    \end{aligned}
    \end{equation}
    Observe that $\psi_\mu^{\text{T}}$ is invariant under \eqref{susygaugetrRS}.
    \item We consider \eqref{transvcond} as a functional condition on the variable 
    \begin{align}
        \psi^u_\mu:=\psi_\mu + \d_\mu u,
    \end{align}
    and solve it explicitly for $u$:
    \begin{equation}
    \label{div-less-dressing}
    \d^{\,\mu} \psi^u_\mu =  \d^{\,\mu} (\psi_\mu + \d_\mu u) = 0 , \quad \Rightarrow \quad u[\psi] = - \Box\- (\d^{\, \mu} \psi_\mu) = - \Box\- (\d^{\, \mu} \psi^\text{L}_\mu) = - \kappa .
    \end{equation}
    \item We check the transformation of $u$. As a functional of $\psi$, under \eqref{psiTkappagaugetr} $u$ gauge transforms as
    \begin{align}
    \label{udressfieldtrepsRS2}
    u[\psi]^\upepsilon \defeq u[\psi^\upepsilon] 
    =  - \kappa^\upepsilon 
    =- \kappa - \upepsilon
    =u[\psi] - \upepsilon,
    \end{align}
    which is the Abelian (additive) version of a \emph{dressing field} transformation \eqref{GT-dressing}. Therefore, we conclude that solving explicitly the divergencelessness condition \eqref{transvcond} does not result in a gauge-fixing, but in a dressing operation via the DFM. This dressing is \emph{non-local}.
    \item The corresponding dressed field reads
    \begin{align}
    \label{div-dressed-field}
    \psi^u_\mu := \psi_\mu + \d_\mu u = 
    \big( \psi^\text{T}_\mu + \d_\mu \kappa \big) -\d_\mu \kappa =
    \psi^{\text{T}}_\mu .
    \end{align}
    It is both susy-invariant and  divergence-free by construction.
    \item In fact, applying the decomposition \eqref{TLdivergdec} to $\psi^u_\mu$, we find
    \begin{equation}
    \begin{aligned}
    \kappa^u & = \kappa + u \equiv 0, \\
    \big(\psi^u_\mu\big)^\text{T} & = \psi^{\text{T}}_\mu=\psi_\mu^u.
    \end{aligned}
    \end{equation}
\end{enumerate}
The dressing field in this case is given in terms of $\kappa$, which carries 4 off-shell d.o.f.; the resulting \emph{dressed field} $\psi^u =\psi_\mu^\text{T}$ 
carries 12 physical off-shell d.o.f. and still contains both a spin-$\sfrac{3}{2}$ (8 d.o.f.) component and a spin-$\sfrac{1}{2}$ (4 d.o.f.) part. Indeed, we may observe that the dressed field is still gamma-traceful, $\gamma^{\,\mu} \psi^u_\mu \neq 0$. The 12 physical off-shell d.o.f. are obtained in a \emph{gauge-invariant} way, without any restriction on the gauge group.\footnote{The functional constraint \eqref{transvcond} is analogous to the  ``\emph{Lorenz gauge}" for a vector gauge potential $A_{\mu}$ in QED: $\d^{\,\mu} A_{\mu}=0$. We refer the reader to \cite{Berghofer-Francois2024} for the proof that, solving explicitly the Lorenz constraint, one ends up building a (non-local) dressing field $u[A]$, which allows to write down the $\U(1)$-invariant (self-)dressed field $A^u_\mu\defeq A_\mu+ \d_\mu u[A]$.}

\smallskip
In both the cases reviewed above, the dressed field $\psi^u$ is a \emph{relational variable} \cite{JTF-Ravera2024c}: it encodes the physical, invariant relations among the off-shell d.o.f. of $\psi$. 
Observe that $\E$-invariance is achieved at the ``cost" of locality, hinting at the fact that susy is, in this context, what is called a \emph{substantial} (or \emph{substantive}) symmetry \cite{Francois2018}.\footnote{As opposed to  \emph{artificial} gauge/local symmetries \cite{Francois2018}, which can be eliminated (reduced) without losing the locality of the theory.}

\paragraph{Dynamics of the dressed theory}

Implementing the DFM at the level of the dynamics in both the cases described above, 
the Lagrangian $4$-form of the dressed theory is, by \eqref{dressed-Lagrangian},
\begin{equation}
\begin{aligned}
\label{dressed-RSLagr}
    L_{\text{RS}}(\psi^u) &= \bar \psi^u \wedge \gamma_5 \gamma \wedge d \psi^u 
     \\
    &=L_{\text{RS}}(\psi) + db(\psi; u) \\
    &=\bar \psi \wedge \gamma_5 \gamma \wedge d \psi + d(\bar u \wedge \gamma_5 \gamma \wedge d \psi ).
\end{aligned}
\end{equation}
In components, we have the dressed Lagrangian density
\begin{align}
    \L_{\text{RS}}(\psi^u) = \epsilon^{\,\mu \nu \rho \sigma} \bar \psi_\mu^u \gamma_5 \gamma_\nu \d_\rho \psi^u_\sigma\,.
\end{align}
The dressed field equations read
\begin{align}
\label{dressed-RSfieldeqs}
    \gamma_5 \gamma \wedge d \psi^u = 0 \quad \rarrow \quad
    \epsilon^{\,\mu \nu \rho \sigma} \gamma_5 \gamma_\nu \d_\rho \psi_\sigma^u = 0.
\end{align}
Let us stress that the dressed Lagrangian \eqref{dressed-RSLagr} is $\E$-invariant because $\psi^u$ is a \emph{susy singlet} ($\E$-invariant). 
Given the relational, gauge-invariant character of the dressed theory, the dressed field equations \eqref{dressed-RSfieldeqs} are deterministic, meaning that, once initial conditions are specified,
they determine in a unique way the evolution of the relational d.o.f. encoded by $\psi^u$.

Finally, we observe that, if $u[\psi]$ is given by the gamma-trace constraint \eqref{gamma-tr-dressing}, the dressed field equations \eqref{dressed-RSfieldeqs} automatically imply $\d^\mu \psi^u_\mu \equiv 0$. On the other hand, if $u[\psi]$ is given by the divergencelessness constraint \eqref{div-less-dressing}, the dressed field equations \eqref{dressed-RSfieldeqs} yield the weaker condition $\slashed{\d}(\gamma^\mu \psi^u_\mu)\equiv 0$.

\subsubsection*{Dressed BRST formalism for the RS case}

Here we provide the BRST formulation of the above discussion. 
We thus introduce the BRST operator $s_{\text{susy}}$ associated with supersymmetry, and the BRST algebra for the RS field,
\begin{equation}
\begin{aligned}
    & s_{\text{susy}} \psi = - dc , \\
    & s_{\text{susy}} c = 0 ,
\end{aligned}
\end{equation}
where $c$ is a spinorial ghost field and its BRST transformation identically vanished due to the additive Abelian character of the symmetry group.
This BRST algebra simply encodes the susy transformations \eqref{susygaugetrRS}. 

For both for the gamma-trace dressing and for the transverse dressing,  we have 
\begin{align}
    s_{\text{susy}} u = -c , 
\end{align}
the BRST version of \eqref{udressfieldtrepsRS1} and \eqref{udressfieldtrepsRS2}. Correspondingly, the dressed (Abelian) ghost is  
\begin{align}
    c^u= c + s_{\text{susy}} u = c -c =0,
\end{align}
which implies the triviality of the dressed BRST algebra:
\begin{equation}
\begin{aligned}
    & s_{\text{susy}} \psi^u = - dc^u = 0 , \\
    & s_{\text{susy}} c^u = 0 ,
\end{aligned}
\end{equation}
i.e. the susy-invariance of $\psi^u$, as expected.  
One can explicitly verify the BRST invariance of $\psi^u$ in both the gamma-trace dressing and the transverse dressing. In the former case, indeed, we have, in $d$ spacetime dimensions,
\begin{align}
    u = u[\psi] = - d \slashed{\d}\- \chi \quad \Rightarrow \quad s_{\text{susy}} \psi^u_\mu = s_{\text{susy}} \left( \psi_\mu - d \partial_\mu \slashed{\d}\- \chi \right) = s_{\text{susy}} \psi_\mu - d \partial_\mu \slashed{\d}\- \left( s_{\text{susy}} \chi \right) = - \partial_\mu c + \partial_\mu c = 0 ,
\end{align}
where we have used the fact that $s_{\text{susy}} \chi = - \tfrac{1}{d}\slashed{\d} c$, the BRST version of the first line of \eqref{susygaugetrRS}.
Similarly, in the case of the transverse dressing, 
\begin{align}
    u = u[\psi] = - \Box\- (\d^{\, \mu} \psi_\mu) = - \kappa \quad \Rightarrow \quad s_{\text{susy}} \psi^u_\mu = s_{\text{susy}} \left( \psi_\mu + \partial_\mu \kappa \right) = s_{\text{susy}} \psi_\mu + \partial_\mu \left( s_{\text{susy}} \kappa \right) = - \partial_\mu c + \partial_\mu c = 0 ,
\end{align}
simply exploiting $s_{\text{susy}} \kappa = - s_{\text{susy}} u = c$.
Finally,  in both the  cases the dressed Lagrangian $L_{\text{RS}}(\psi^u)$ satisfies  $s_{\text{susy}} L_{\text{RS}}(\psi^u) \equiv 0$, as susy is completely reduced and the dressed Lagrangian written in terms of susy singlets.
In the presence of another gauge group $\J$ besides susy, one should consider $s=s_{\text{susy}} + s_J$, where $s_J$ is the BRST operator associated with the residual $\J$-transformations: after susy reduction, the dressed ghost depends on $c_J$ and $s_J u$, see eq. \eqref{dressed-ghost-res1stkind}, so that the dressed BRST algebra \eqref{dressed-BRSTalgebradef} encodes the linear $\J$-transformation of $\psi^u$.

\subsection{Gravitino self-dressing and relational supergravity}\label{Gravitino self-dressing in supergravity}

We now turn to the case of supergravity (sugra), where things become a bit more subtle  because susy transformations correspond to diffeomorphisms along the fermionic directions of \emph{superspace} $M^{d|\n \mathcal{N}}$.\footnote{We denote the number of spinorial dimensions by $\n$, which, in $d=4$ spacetime dimensions, is $\n=2^{d/2}=4$. Recall that superspace is spanned by bosonic and Grassmannian coordinates $\lbrace{x^{\,\mu},\theta^{\,\alpha}\rbrace}$.} 
This is made clearer  in the geometric  approach to supergravity in superspace: See for example the so-called \emph{rheonomic} approach \cite{Neeman-Regge1978,Neeman-Regge1978b,Castellani:1991eu} comprehensively reviewed in  \cite{DAuria:2020guc,Castellani:2019pvh,JTF-Ravera2024review}, or more generally the framework of Cartan super-geometry as the mathematical foundation of sugra -- see  \cite{JTF-Ravera2024review} and references therein. 
In order to have better geometric control over the procedure to be implemented via DFM for reducing susy in sugra (and obtain gauge-invariant field-theoretical variables) one should adopt a super-Cartan approach,  that is, work with a super-Cartan bundle. 
Moreover, since higher-dimensional, $\mathcal{N}$-extended sugra models typically involve higher-degree forms (antisymmetric tensors with multiple indices), a fully developed relational framework for sugra would require the formal development of higher Cartan supergeometry, 
along with a corresponding higher version of the DFM (the latter being in preparation \cite{JTF-Ravera2025higherDFM}).

Here, we will focus on the simple case of  $\mathcal{N}=1$, $d=4$ sugra, suggesting a dressing \emph{superfield} within the rheonomic approach, which is enough to exemplify our relational approach to sugra. 
We shall first review the coupling with gravity and the infinitesimal local susy transformations -- the sugra literature indeed typically deals only with infinitesimal transformations. 
We will see how the gamma-trace and covariant transverse constraints in sugra are obtained via the DFM.

\smallskip
The minimal coupling of the RS field with gravity is
described via the (Lorentz) spin connection ${\omega^a}_b \in \Omega^1\big(U, \so(1,3)\big)$ and the soldering 1-form $e^a \in \Omega^1\big(U, \RR^4\big)$ (the vierbein). It is obtained as usual via the substitution (``covariantization")
\begin{equation}
\begin{aligned}
    \d_\mu & \,\mapsto\, D_\mu \defeq\d_\mu + \rho_*(\omega_\mu) , \\
    \gamma_\mu & \,\mapsto\, \gamma_a {e^a}_\mu ,
\end{aligned}
\end{equation}
where $D_\mu$ is the $\SO(1,3)$-covariant derivative and $\gamma_a$  the flat space gamma-matrices. 
Then, the complete spacetime Lagrangian describing both the dynamics of gravity (spin-$2$ graviton) and of the RS field $\psi$, now renamed ``gravitino", is
\begin{align}
\label{sugraL}
    L_{\text{\tiny{sugra}}}(\omega^{ab},e^a,\psi) = R^{ab} \wedge e^{c} \wedge e^{d} \epsilon_{abcd} + 4 \bar \psi \wedge \gamma_5 \gamma \wedge D \psi ,
\end{align}
where $R^{ab}:=d\omega^{ab}+{\omega^a}_c \wedge \omega^{cb}$ is the Riemann (Lorentz) curvature 2-form
and $\gamma:= \gamma_a e^{a}= \gamma_a {e^{a}}_{\!\mu} \, d x^{\,\mu}:=\gamma_\mu \, d x^{\,\mu}$.

The \emph{infinitesimal transformation} \eqref{susygaugetrRS} is not a symmetry of the coupled theory \eqref{sugraL}. We have to covariantize it,
\begin{align}
\label{sugrasusytr1}
    \delta_\epsilon \psi = D \epsilon \quad \rarrow \quad \delta_\epsilon \psi_\mu = D_\mu \epsilon .
\end{align}
The infinitesimal $\E$-transformations of $e^a$ and ${\omega^a}_b$ are \cite{Castellani:1991eu,Tanii:2014gaa}
\begin{equation}
\begin{aligned}
\label{sugrasusytr2}
    \delta_\epsilon e^a & = i \, \b \epsilon  \, \gamma^a \psi , \\
    \delta_\epsilon \omega^{ab} & = i \, \big(  \bar D^{[a} \bar \psi^{b]} \gamma_c - 2 \, \bar D^{[a} \bar \psi_c \gamma^{b]} \big) \, \epsilon \, e^c .
\end{aligned}
\end{equation}
Under \eqref{sugrasusytr1}-\eqref{sugrasusytr2} the Lagrangian \eqref{sugraL} is quasi-invariant.
The field equations of the theory are
\begin{equation}
\begin{aligned}
     \label{field-eq-sugra}
    \bs E(\omega)& = 2 (De^c - \sfrac{i}{2} \bar \psi \wedge \gamma^c \psi) \wedge e^d \epsilon_{abcd} = 0, \\
     \bs E(e) &= 2 R^{ab} \wedge e^c \epsilon_{abcd} + 4 \bar \psi \wedge \gamma_5 \gamma_d D \psi = 0 , \\
      \bs E(\psi)&= 8 \gamma_5 \gamma_a D \psi \wedge e^a + 4 \gamma_5 \gamma_a \psi \wedge (De^a - \sfrac{i}{2} \bar \psi \wedge \gamma^a \psi)= 0.
\end{aligned}
\end{equation}
We remind that the susy algebra --  the  commutator of two susy transformations with parameters $\varepsilon$ and $\varepsilon'$ -- closes only \emph{on-shell}, i.e. on the field equations. 
Anticipating future comments, we may rephrase by saying that the closure of the susy transformations on \emph{bare} fields occurs only on-shell.

In the sugra literature, to get the right number of off-shell d.o.f. (12 for the gravitino), one usually ``gauge-fixes" $\psi$ by requiring the condition $\gamma^\mu \psi_\mu=0$. Below  we show that solving the functional constraint associated with this ``gauge-choice" actually results in a \emph{perturbative dressing} via the DFM.

\paragraph{Gamma-trace dressing in sugra}

Let us follow the template established for the pure RS case, just ``covariantizing" where needed.
We consider the gamma-tracelessness constraint \eqref{gammatr} as a functional condition on the variable 
\begin{align}
    \psi_\mu^\upsilon \defeq \psi_\mu+ \b\updelta_\upsilon \psi ,
\end{align}
and solve it explicitly for the linear ``parameter" $\upsilon$:
\begin{equation}
\label{gamma-tr-dressing-sugra}
    \gamma^{\,\mu} \psi^\upsilon_\mu = \gamma^{\,\mu} (\psi_\mu + D_\mu \upsilon) = 0 \quad \Rightarrow \quad \upsilon[\psi] = - \slashed{D}\- (\gamma^{\,\mu} \psi_\mu) = - d \slashed{D}\- \chi.
\end{equation}
Then, we check that $\upsilon[\psi]$ satisfies \eqref{pert-dressing-field}, neglecting higher-order terms. Indeed,  
given $\delta_\epsilon \, \chi = \frac{1}{d} \slashed{D} \epsilon$, one has
\begin{align}
    \delta_\epsilon \upsilon[\psi]= 
     - d \slashed{D}\- (\delta_\epsilon \, \chi ) 
     - d \cancelto{\,\text{\tiny{neglect}}}{\delta_\epsilon  (\slashed{D}\-) \chi}\!\!\!\!\!
     \approx - \epsilon.
\end{align}
So $\upsilon[\psi]$ is indeed a perturbative dressing field.
We therefore build the \emph{perturbatively dressed gravitino} field
\begin{align}
    \psi^\upsilon \defeq \psi + D\upsilon[\psi] 
    = \psi - d D \slashed{D}\- \chi,
\end{align}
which, by construction, fulfills $\gamma^{\,\mu} \psi_\mu^\upsilon\equiv0$ and is susy-invariant at 1st order, $\delta_\epsilon \psi^\upsilon \approx 0$.
Hence, we can immediately conclude that what is usually referred to as the gravitino field is, in fact, a \emph{susy-invariant self-dressed} \emph{non-local} field carrying 12 (\emph{relational}) d.o.f. off-shell.

\paragraph{Covariant transverse dressing}

Analogously, one can consider the functional constraint \eqref{transvcond} on the variable
\begin{align}
    \psi_\mu^\upsilon \defeq \psi_\mu+ \b\updelta_\upsilon \psi
\end{align}
and solve it explicitly for $\upsilon$:
\begin{equation}
\label{div-dressing-sugra}
    D^{\,\mu} \psi^\upsilon_\mu = D^{\,\mu} (\psi_\mu + D_\mu \upsilon) = 0 \quad \Rightarrow \quad \upsilon[\psi] = - \Box\- (D^{\,\mu}\psi_\mu) .
\end{equation}
It is easily checked -- again neglecting higher-order terms -- that 
\begin{align}
    \delta_\epsilon \upsilon[\psi] =  -  \Box\- (D^{\,\mu} \delta_\epsilon(\psi_\mu)) - \cancelto{\,\text{\tiny{neglect}}}{\delta_\epsilon (\Box\- D^{\,\mu}) \psi_\mu} \!\!\!\!\!\approx -\epsilon.
\end{align}
We can then write down the perturbatively dressed, \emph{non-local} gravitino,
\begin{align}
    \psi^\upsilon := \psi + D\upsilon[\psi] = \psi - D[\Box\- (D^{\,\mu}\psi_\mu)] .
\end{align}
By construction, it is divergence-free, $D^{\,\mu} \psi_\mu^\upsilon\equiv0$, and susy-invariant at 1st order, $\delta_\epsilon \psi^\upsilon \approx 0$.
The dressed gravitino $\psi^\upsilon$ is transverse, $\psi^\upsilon =\psi^\text{T}$, and carries 12 off-shell d.o.f. -- as it can be easily checked by considering the covariant version of \eqref{TLdivergdec}, that is $\psi=\psi^\text{T}+\psi^\text{L}=\psi^\text{T}+D_\mu[\Box\-(D^{\,\nu}\psi_\nu)]$.

\paragraph{Dressed supergravity and its dynamics}

In both the cases discussed above, the perturbatively dressed vielbein and spin connection 1-forms are formally given by
\begin{equation}
\begin{aligned}
\label{dressed-vielbeinandspinconn}
    (e^{a})^\upsilon &:= e^a + i \, \b \upsilon[\psi]  \, \gamma^a \psi , \\
    (\omega^{ab})^\upsilon &:= \omega^{ab} + i \, \big(  \bar D^{[a} \bar \psi^{b]} \gamma_c - 2 \, \bar D^{[a} \bar \psi_c \gamma^{b]} \big) \, \upsilon[\psi] \, e^c .
\end{aligned}
\end{equation}
These perturbatively dressed fields are \emph{relational variables} \cite{JTF-Ravera2024c}: they represent the physical, invariant relations among the (off-shell) d.o.f. of $\omega$, $e$, and $\psi$. 

According to the DFM -- see 
\eqref{pert-dressed-Lagrangian} -- with either dressing fields above the Lagrangian $4$-form of the dressed theory is written as
\begin{align}
\label{pert-dressed-SUGRALagrangian}
  L_{\text{\tiny{sugra}}}(\omega^\upsilon, e^\upsilon, \psi^\upsilon)\defeq L_{\text{\tiny{sugra}}}(\omega, e, \psi) + d\beta(\omega, e, \psi; \upsilon). 
\end{align}
Let us stress that it is susy-invariant at 1st order because $\psi^\upsilon$, $\omega^\upsilon$, and $e^\upsilon$ are  susy \emph{singlets} at 1st order.
The dressed field equations read
\begin{equation}
\begin{aligned}
     \label{dressed-SUGRAfield-eqs}
    \bs E(\omega^\upsilon)& = 2 [D^\upsilon(e^c)^\upsilon - \sfrac{i}{2} \bar \psi^\upsilon \wedge \gamma^c \psi^\upsilon] \wedge (e^d)^\upsilon \epsilon_{abcd} = 0, \\
     \bs E(e^\upsilon) &= 2 (R^{ab})^\upsilon \wedge (e^c)^\upsilon \epsilon_{abcd} + 4 \bar \psi^\upsilon \wedge \gamma_5 \gamma_d D^\upsilon \psi^\upsilon = 0 , \\
      \bs E(\psi^\upsilon)&= 8 \gamma_5 \gamma_a D^\upsilon \psi^\upsilon \wedge (e^a)^\upsilon + 4 \gamma_5 \gamma_a \psi^\upsilon \wedge [D^\upsilon(e^a)^\upsilon - \sfrac{i}{2} \bar \psi^\upsilon \wedge \gamma^a \psi^\upsilon]= 0,
\end{aligned}
\end{equation}
and they are deterministic, meaning that they uniquely determine the evolution of the relational d.o.f. of the theory.
This showcases the simplest model of \emph{relational supergravity}, where susy-invariance is implemented at 1st order.

\paragraph{Dressing superfield in superspace}

In \cite{JTF-Ravera2024-SUSY} a candidate for a possible \emph{dressing superfield} within the \emph{rheonomic approach} \cite{Castellani:1991eu} to supergravity in superspace was proposed. This was done by considering the extension of the functional constraint \eqref{gammatr} to superfields in superspace. We briefly review this construction. 
The gravitino super 1-form field in superspace is
\begin{align}
\label{superfieldeq}
    \psi^{\,\alpha}(x,\theta) = {\psi^{\,\alpha}}_\mu (x,\theta) \,d x^{\,\mu} + {\psi^{\,\alpha}}_\beta (x,\theta)\, d \theta^{\,\beta} .
\end{align}
The infinitesimal transformation of $\psi^{\,\alpha}(x,\theta)$ under (even) superdiffeomorphisms in $M^{4|4}$ is given by the Lie derivative $\ell_X \psi:= d(\iota_X \psi) + \iota_X (d\psi) $ along a vector superfield $X$.
In the $\mathcal{N}=1$ susy case, we have
\begin{align}
\label{superspace-susytr}
    \ell_X \psi = D \epsilon =: \delta_\epsilon \psi ,
\end{align}
where $\epsilon=\epsilon^{\,\alpha}=\epsilon^{\,\alpha}(x,\theta)=\iota_X \psi^{\,\alpha}=X^{\,\mu} {\psi^{\,\alpha}}_\mu + X^{\,\beta} {\psi^{\,\alpha}}_\beta$ is the local susy (super)parameter. 
One then considers the following ``gamma-trace decomposition" of $\psi^{\,\alpha}(x,\theta)$:
\begin{equation}
\begin{aligned}
\label{superdec}
    \psi^{\,\alpha}(x,\theta) & = \big[{\uprho^{\,\alpha}}_{\,\mu}(x,\theta)+ {(\gamma_\mu)^{\,\alpha}}_\beta \, \chi^{\,\beta} (x,\theta) \big] \, dx^{\,\mu} + \big[ {\uprho^{\,\alpha}}_{\,\beta}(x,\theta)+ {(\gamma_\mu)^{\,\alpha}}_\beta \, \chi^{\,\mu} (x,\theta) \big] \, d\theta^{\,\beta} \\
    & = {\uprho^{\,\alpha}}_{\,\mu}(x,\theta) \, dx^{\,\mu} + {\uprho^{\,\alpha}}_{\,\beta}(x,\theta) \, d\theta^{\,\beta} + {(\gamma_\mu)^{\,\alpha}}_\beta \big[ \,\chi^{\,\beta} (x,\theta) \, dx^{\,\mu}  +  \chi^{\,\mu} (x,\theta) \, d\theta^{\,\beta}\,  \big] ,
\end{aligned}
\end{equation}
with
\begin{align}
    {(\gamma^{\,\mu})^\alpha}_\beta \,{\uprho^{\,\beta}}_\mu(x,\theta)=0, \qquad {(\gamma_{\nu})^\alpha}_\beta \,{\uprho^{\,\beta}}_\alpha(x,\theta)=0 .
\end{align}
Notice that $\chi^{\,\beta} (x,\theta)$ is a spin-$\sfrac{1}{2}$ superfield, while $\chi^{\,\mu} (x,\theta)$ is a 4-vector (spin-$1$) superfield; they are defined as
\begin{equation}
\begin{aligned}
\label{superdef}
\chi^{\,\alpha} (x,\theta) &:= \frac{1}{d} \,{(\gamma^{\,\mu})^{\,\alpha}}_\beta \, {\psi^{\,\beta}}_\mu (x,\theta) , \\
\chi^{\,\mu} (x,\theta) &:= \frac{1}{\n} \,{(\gamma^{\,\mu})^{\,\alpha}}_\beta \,{\psi^{\,\beta}}_\alpha (x,\theta) ,
\end{aligned}
\end{equation}
where $\n=2^{d/2}$ is the number of spinorial dimensions -- i.e., here in $d=4$, $\n=4$. 
Now, considering the gamma-tracelessness constraint
\begin{align}
    \gamma \cdot \psi := {(\gamma^{\,\mu})^{\, \alpha}}_\beta \, {\psi^{\,\beta}}_Z (x,\theta)=0, \quad Z = \lbrace{\alpha,\mu \rbrace},
\end{align}
on the 1-form superfield $\psi^\alpha$ in \eqref{superfieldeq} as a functional condition on the ``super-variable" 
\begin{align}
    \psi^\upsilon:=\psi + \b\updelta_\upsilon \psi
\end{align}
and solving it explicitly for the linear ``super-parameter" $\upsilon$, one gets
\begin{equation}
\label{superspace-gamma-tr-dressing-sugra}
    \gamma \cdot \psi^\upsilon = \gamma \cdot (\psi + D \upsilon) = 0 , \quad
    \Rightarrow \quad \upsilon[\psi] = - \slashed{D}\- (\gamma \cdot \psi) .
\end{equation}
We can then easily check that $\upsilon[\psi]$ is actually a \emph{dressing superfield} in superspace: it satisfies, neglecting higher-order terms,
\begin{align}
    \delta_\epsilon \upsilon[\psi]= 
    - \slashed{D}\- (\gamma \cdot \delta_\epsilon \psi) 
     \approx - \epsilon,
\end{align}
fulfilling the defining property \eqref{pert-dressing-field}.
Finally, the perturbatively dressed gravitino 1-form superfield, which is a \emph{self-dressed superfield}, reads
\begin{align}
    \psi^\upsilon \defeq \psi + D\upsilon[\psi] 
    = \psi - D \slashed{D}\- (\gamma \cdot \psi) .
\end{align}
By construction, it is gamma-traceless, $\gamma\cdot \psi\equiv0$, and susy-invariant at 1st order, $\delta_\epsilon \psi^\upsilon \approx 0$.
Following the prescription given by the rheonomic approach, we may then restrict the theory to spacetime, which yields, 
\begin{align}
    \psi^{\,\alpha}(x,\theta)|_{\theta=d\theta=0}=\big[  {\rho^{\,\alpha}}_\mu (x) + {(\gamma_\mu)^{\,\alpha}}_\beta \, \chi^{\,\beta} (x) \big]\, dx^{\,\mu}.
\end{align}
We recover the previous results  for the dressed sugra theory. For the dressed gravitino 1-form in fact we simply have $\psi^\upsilon(x,\theta)|_{\theta=d\theta=0}=\psi^\upsilon(x)$, and similarly for the other fields of the theory.

\section{Matter-Interaction Supergeometric Framework}
\label{Matter-Interaction Supergeometric Framework}

In this section we will 
{articulate the template of}
a novel Matter-Interaction Supergeometric Unification (MISU) scheme, {inspired by the analysis of ``unconventional supersymmetry" (ususy) via the DFM done in \cite{JTF-Ravera2024ususyDFM}. 
We shall illustrate this general template by treating the simplest MISU kinematics based on the Lorentz superalgebra, and deriving its dressed BRST algebra. 
Finally, we comment on how the ususy notion is a special case of the proposed MISU framework, stressing the root cause of previously unsuccessful attempts to extend ususy beyond the model in which it was showcased.
}

\subsection{Relational Matter-Interaction Supergeometric Unification via the DFM}\label{Matter-Interaction Supergeometric Unification via the DFM}

Supersymmetric field theory as conventionally applied within high energy physics implies a doubling of the number of fundamental fields (and particles), whereby each known particle has a \emph{superpartner} of opposite statistics: fermionic (matter) fields have bosonic partners, bosonic (gauge and Higgs) fields have fermionic partners. 
This view can only be accommodated with empirical data via
the notion of (spontaneous) susy breaking.
Still, until now, supersymmetric particles did not show up in colliders. 

However, this fact does not decisively undermine the framework of differential supergeometry (the mathematical foundation of supersymmetric field theory), which \emph{does not} require a matching of bosonic and fermionic degrees of freedom (d.o.f.) \cite{Sohnius:1985qm}.
The MISU framework {we propose here}  freely exploits supergeometry without the d.o.f. matching constraint to describe gauge interactions and matter fields as {\emph{parts of the same superconnection}.} 
It is thus closer to Berezin's original motivation for the introduction of supergeometry in fundamental
physics \cite{Berezin-Marinov1977}:
a program currently ``in dormancy" that  awaits to yield its full potential, and of which our proposal is but the first step. 
Furthermore, the DFM is an integral part of MISU, so that the {unifying} invariant superconnection is fundamentally a \emph{relational variable}.
{Let us sketch the template for a MISU kinematics.}

\medskip

In our Matter-Interaction Supergeometric Framework, 
the kinematics is given by
a superbundle 
{$P\rarrow M$ where the  base $M$ is bosonic and whose structure group $H$ is a graded Lie group: correspondingly, the gauge super-group is $\H$, i.e. susy is ``\emph{internal}". 
So that the full super-group of local transformations is $\Diff(M)\ltimes \H$.
The field space on which it acts is $\Phi=\{e, \mathbb{A}\}$ where $e=e^a={e^a}_\mu\, dx^{\,\mu}$
  is the canonical soldering form of $M$ and  $\mathbb{A}$ is a (non-canonical) $\LieH$-valued Ehresmann superconnection, $\LieH$ the Lie superalgebra of $H$.}
  


{There are two options for choosing  $\LieH$ in a MISU kinematics.
First and easiest, one may simply select the Lie superalgebra  $\LieH$  
 s.t. the bracket of its susy generators $\mathbb Q$ is not generated by  ``(internal) infinitesimal translation generators" $\mathbb P$.
 This is so that there is no spurious ``internal translation potential" in $\mathbb{A}$, which would be redundant (or would need to be identified) with the existing soldering $e$ of $M$.}
 
{Secondly, less trivially, one may} consider  Cartan super-geometries \cite{JTF-Ravera2024review,JTF-Ravera2024ususyDFM} modeled on  Klein pairs of Lie superalgebras $(\mathfrak{g}, \mathfrak{h})$  s.t. $\mathfrak{h}\subset \mathfrak{g}$ 
and $\mathfrak{g}/\mathfrak{h}$ is a bosonic {$H$-module} with basis (generators) $\mathbb P$ (infinitesimal ``translations"): 
{Then, $P\rarrow M$ is a superbundle as required above, and we have a Cartan geometry  $(P, \b{\mathbb{A}})$ with
 $\LieG$-valued Cartan super-connection $\b{\mathbb{A}}$  spliting as a $\LieH$-valued Ehresmann super-connection $\mathbb{A}$ and a  $\mathfrak{g}/\mathfrak{h}$-valued soldering form (vielbein) $e=e^a$; i.e. the field space is the space of Cartan connections $\Phi=\{\b{\mathbb{A}} \}$, which is naturally subject to internal transformations under 
 $\Diff(M)\ltimes \H$
 (which contains no ``gauged translations").  
 } 

{To find  candidates suiting either of the above desiderata}, for $\LieH$ and/or $(\LieG, \LieH)$, one can refer to, and pick from, a classification of superalgebras, e.g. \cite{Kac1977}.

{In either of the two described options, to obtain the  MISU kinematics one must finally apply the DFM:} 
In both cases, the susy {($\mathbb Q$-)} part of the superconnection $\mathbb A$ is indeed a spinor 1-form field $\psi=\psi^\alpha$, from which a susy-dressing field $u[\mathbb A]$ can be extracted via a gamma-trace decomposition as illustrated {in Sections \ref{Rarita-Schwinger spinor-vector self-dressing} and \ref{Gravitino self-dressing in supergravity}.} 
{Using~\eqref{dressed-fields} (or \eqref{pert-dressed-fields}) of Section \ref{The Dressing Field Method}}, one  obtains a susy-invariant dressed superconnection $\mathbb{ A}^u$ whose susy component is of the form $\gamma \, \chi^u = dx^{\,\mu}\gamma_\mu \,\chi^u$ with $\chi^u$ a spinor field, {potentially describing a \emph{matter field}} -- and $\gamma=dx^{\,\mu}\gamma_\mu = dx^{\,\mu}\gamma_a {e^a}_\mu$ the gamma matrix 1-form --
{and whose bosonic component 
describes a gauge potential.}
Thus, the  susy-invariant relational superconnection $\mathbb A^u$ geometrically (kinematically) unifies gauge and matter fields. 
\medskip

{Given a MISU kinematics obtained following the above template, many models may be built, i.e. a dynamics maybe chosen by writing down a Lagrangian. 
 Model building can be approached in two ways.}
 
{One option is to propose a
 Lagrangian that is (quasi-)invariant under either the initial full  supergroup $\H$ -- that is, covariant under $\Diff(M) \ltimes \H$ -- i.e. to apply the Gauge Principle to the bare (``pre-MISU") kinematics, and then dress it via 
 \eqref{dressed-Lagrangian} (or \eqref{pert-dressed-Lagrangian}). 
 One thereby gets a dressed Lagrangian, from which derive dressed field equations for the dressed fields, that is a MISU model.
 In Section \ref{The AVZ model as a MISU subcase} we discuss such a case, the so-called AVZ model of ususy.}
 
 {Alternatively, one may start  directly from the MISU kinematics, whose residual gauge transformations of the 1st kind (remind Section \ref{The Dressing Field Method}) is the bosonic gauge group $\H/$susy, i.e. whose group of local transformations is $\Diff(M) \ltimes (\H/$susy$)$. 
 That is, one may apply the Gauge Principle to the residual gauge transformations of the 1st kind,  proposing a Lagrangian (quasi-)invariant under $\H/$susy, whose field equations for the dressed field are therefore ($\H/$susy)-covariant.}
 {The first approach is more constraining than the second, the latter's gauge group being only a subgroup of the former's.\footnote{We observe that the second approach is that occurring in Poincaré and Metric-Affine gravity, where the ``gauge translations" are discarded or reduced via dressing (as shown in \cite{JTF-Ravera-2025-MAG}) \emph{before} model model building starts, being thus constrained only by (residual) invariance under the (residual) Lorentz $\SO(1,3)$ or General Linear $\GL(n)$ groups, rather than by the gauged Poincaré or affine groups.}
We illustrate next our template with a simple MISU kinematics. }

\subsection{{An illustrative model: Lorentz superalgebra-based MISU}}\label{An illustrative model: Lorentz superalgebra-based MISU}

{Let us consider} a simple example in {$n=4$}  spacetime dimensions: the semi-direct extension $\LieH=\mathfrak{sl}(2, \CC) \oplus \CC^2$ of the Lorentz algebra $\mathfrak{spin}(1,3)\simeq\mathfrak{sl}(2, \CC)$ by an (additive) Abelian superalgebra. 
We {consider} a superbundle {$P\rarrow M$} with structure supergroup $H=S\!L(2, \CC) \ltimes \CC^2$ over an even manifold $M$, with {corresponding} gauge supergroup $\H:= \{ g, g':  M \rarrow H\,|\, {g}^{\prime\, g} =g\- g' g \}$ and  gauge algebra
Lie$\H:= \{ \lambda, \lambda': M \rarrow~\LieH\,|\, \delta_\lambda \lambda' = [\lambda', \lambda] \}$. 
Relying on a compact matrix notation, we write an element of $\H$ as
\begin{align}
    g=\begin{pmatrix} B & \upepsilon\\ 0 & 1 \end{pmatrix},
\end{align}
with $B\in \SL(2, \CC)$ and $\upepsilon:M \rarrow  \CC^2$ a susy spinor, and the semi-direct structure of $\H$ is reproduced by matrix multiplication.
Let us introduce the notation 
\begin{align}
  \mathbb{B} \defeq \begin{pmatrix} B & 0\\ 0 & 1 \end{pmatrix} , \quad \quad \boldsymbol{\upepsilon} \defeq  \begin{pmatrix} \1 & \upepsilon \\ 0 & 1 \end{pmatrix} 
\end{align}
for the $\SL(2, \CC)$ and the susy gauge subgroups elements, respectively.
Correspondingly, an element of Lie$\H$ is
\begin{align}
\label{gaugeparam}
\lambda 
&= \begin{pmatrix}
   \beta  & 
 \varepsilon \\
0  &\    0
    \end{pmatrix} ,
\end{align}
with $\beta \in \mathfrak{sl}(2, \CC)$ and $\varepsilon \in \CC^2$ a spinorial susy parameter. 
The Ehresmann superconnection and its curvature are thus
\begin{equation}
\begin{aligned}
\label{connection-ex1}
    \mathbb{A} 
= \begin{pmatrix}
  \omega &\   \psi \\
   0  & \  0
\end{pmatrix},
\quad \text{and} \quad\ 
\mathbb{F} =d \mathbb{A}  + \tfrac{1}{2}[\mathbb{A} , \mathbb{A} ] = d \mathbb{A} + \mathbb{A}^2
= \begin{pmatrix}
  \Omega &\   \nabla \psi \\
   0  & \  0
\end{pmatrix}
=\begin{pmatrix}
  d\omega+\omega^2 &\   d\psi+ \omega \psi \\
   0  & \  0
\end{pmatrix},
\end{aligned}
\end{equation}
where  $\nabla$ denotes the Lorentz-covariant derivative.
The curvature satisfies Bianchi identity $D^\mathbb{A}\mathbb{F}= d\mathbb{A} +[\mathbb{A}, \mathbb{F}]=0$.
The $\H$-transformations of $\mathbb{A}$ and $\mathbb{F}$ are
\begin{equation}
\label{GT-A-F-MISU}
\begin{aligned}
\mathbb A^g&=g\- \mathbb A g +g\-dg 
 = \begin{pmatrix}
       B\-\omega B + B\- dB & B\- (\psi+  \nabla \upepsilon ) \\ 0 & 0 
  \end{pmatrix} , \\[2mm]
   \mathbb F^{\,g}&=g\- \mathbb F g 
   = \begin{pmatrix}
       B\-\Omega B  & B\- (\nabla \psi +   \Omega\upepsilon ) \\ 0 & 0 
   \end{pmatrix},
\end{aligned}
\end{equation}
That is, we have for our elementary fields $\omega$ and $\psi$,
\begin{equation}
\begin{aligned}
    \omega^B &= B^{-1} \omega B + B^{-1} dB, \quad &&\text{and} \quad \  
    \omega^\upepsilon = \omega , \\
    \psi^B &= B^{-1}  \psi , \quad &&\text{and} \quad\ 
    \psi^\upepsilon = \psi + \nabla \upepsilon .
\end{aligned}
\end{equation}
Notice that here $\omega$ is a susy singlet already.
The corresponding Lie$\H$-transformations  are
\begin{equation}
\label{linear-GT-A-F}
\begin{aligned}
    \delta_\lambda \mathbb A = D^\mathbb{A}\lambda
    =\begin{pmatrix}
      \nabla \beta  & \ -\beta \psi+ \nabla \varepsilon  \\ 0 & 0 
  \end{pmatrix},
  \quad \text{ and } \quad 
  \delta_\lambda \mathbb F = [\mathbb{F}, \lambda]
     =\begin{pmatrix}
      \left[\Omega, \beta\right]  & \ -\beta \nabla \psi + \Omega \varepsilon  \\ 0 & 0 
  \end{pmatrix},
\end{aligned}
\end{equation}
with $\nabla \beta = d\beta + [\omega ,\beta]$ and $\nabla \varepsilon = d\varepsilon + \omega \varepsilon$.
Again, notice that $\delta_\varepsilon \omega = 0$.

One may now  introduce the  ``gamma-trace" decomposition for $\psi=\psi_\mu \,dx^{\mu}$ precisely as in \eqref{gammatracedec}:  
$\psi =\uprho + \gamma \chi$, 
where by definition $\gamma^{\,\mu} \uprho_\mu=0$ and $\chi:= \frac{1}{4} \gamma^{\,\mu} \psi_\mu$. From it we will extract of a susy dressing field $u=u[\psi]$.
Under (finite) susy transformations we have: 
\begin{align}
    \psi^\upepsilon_\mu = \uprho^\upepsilon_\mu + \gamma_\mu \,\chi^\upepsilon
    \quad 
    \Rightarrow 
    \quad
    \left\{ 
    \begin{matrix}
     \ \ \uprho^\upepsilon_\mu = \uprho_\mu - \tfrac{1}{4} \gamma_\mu \slashed \nabla \upepsilon + \nabla_\mu \upepsilon ,
     \\[2mm]
     \hspace{-12mm} \chi^\upepsilon= \chi + \tfrac{1}{4} \slashed\nabla \upepsilon.
    \end{matrix}
    \right.
\end{align}
Let us define the (field-dependent) operator 
\begin{align}
    b_\mu =b_\mu[\omega] \defeq \tfrac{1}{4} \gamma_\mu \slashed \nabla  - \nabla_\mu  ,
\end{align}
with formal left inverse is $[b\-]^\mu$,
so that we may write
\begin{align}
    \uprho^\upepsilon_\mu = \uprho_\mu - b_\mu (\upepsilon).
\end{align}
Now, given a susy-valued parameter $u$, we  consider 
 the  variable
\begin{align}
    \psi^u_\mu \defeq \psi_\mu + \nabla_\mu u= \uprho^u_\mu + \gamma_\mu \,\chi^u , 
    \quad 
    \text{satisfying the functional constraint }
        \uprho^u_\mu = 0. 
\end{align}
 By solving explicitly for $u$ we extract the \emph{field-dependent dressing field}:
\begin{align}
\label{drfieldmisu}
    u[\psi, \omega]=u[\mathbb A] \defeq [b^{-1}]^{\,\mu} (\uprho_\mu),
\end{align}
whose  dependence on $\omega$ comes from $b_\mu$.
Its susy transformation is indeed easily found to be
\begin{align}
\label{Abelian-misu-dressing}
    u^\upepsilon = u[\mathbb A^\upepsilon] = u[\psi^\upepsilon,\omega^\upepsilon] = u[\psi^\upepsilon,\omega] = (b[\omega]^{-1})^{\,\mu} (\uprho^\upepsilon_\mu) = (b[\omega]^{-1})^{\,\mu} (\uprho_\mu - b_\mu (\upepsilon)) = u[\mathbb A] - \upepsilon.
\end{align}
We have thus a  non-local Abelian dressing field.
We use the matrix notation,
\begin{align}
\boldsymbol{u}=
\begin{pmatrix}
\1 & \  u \\
0 & 1
\end{pmatrix}=\begin{pmatrix}
\1 & \  u[\mathbb A] \\
0 & 1
\end{pmatrix},
\end{align}
so that the standard (``non-Abelian") defining property of a dressing fields is
\begin{align}
\label{susy-GT-u}
\boldsymbol{u}^{\boldsymbol{\upepsilon}} = {\boldsymbol{\upepsilon}}^{-1} \boldsymbol{u} =\begin{pmatrix}
\1 & \  -\upepsilon \\
0 & 1
\end{pmatrix}
\begin{pmatrix}
\1 & \  u \\
0 & 1
\end{pmatrix}= \begin{pmatrix}
\1 & \  u - \upepsilon \\
0 & 1
\end{pmatrix},
\end{align}
reproducing the Abelian transformation \eqref{Abelian-misu-dressing}. 
With the dressing field $\boldsymbol{u}$ we build the susy-invariant  superconnection
\begin{align}
\label{dressed-A}
\mathbb{A}^{\boldsymbol{u}} \defeq \boldsymbol{u}\- \mathbb{A} \boldsymbol{u} + \boldsymbol{u}\- \, d \boldsymbol{u} = \begin{pmatrix}
 \ \omega^u &\  \psi^u + \nabla u \\
   0  &\ 0
\end{pmatrix} = \begin{pmatrix}
 \ \omega &\  \psi + \nabla u \\
   0  &\ 0
\end{pmatrix}
=
\begin{pmatrix}
 \ \omega &\  \gamma \, \chi^u \\
   0  &\ 0
\end{pmatrix} ,
\end{align}
with  the susy-invariant dressed spinor $\chi^u \defeq \chi + \tfrac{1}{4} \slashed{\nabla} u$, by construction. 
Thus, gauge and matter fields feature on equal footing in the dressed superconnection $\mathbb A^{\boldsymbol{u}}$. 
The corresponding dressed curvature is
\begin{align}
\label{dressed-F}
\mathbb{F}^{\,\boldsymbol{u}} = \begin{pmatrix}
     \Omega^u &\  \gamma_a T^a \chi^u - \gamma \,\nabla \chi^u \\ 
     0 & 0
\end{pmatrix} = \begin{pmatrix}
     \Omega &\ \gamma_a T^a \chi^u - 
 \gamma \, \nabla \chi^u \\ 
     0 & 0
\end{pmatrix},
\end{align}
where $T^a\defeq de^a + {\omega^a}_b e^b$ is the torsion of $M$.

Let us remark that  the dressing field $u$ {(or, in the compact matrix notation, $\boldsymbol{u}$)} being non-local -- {by eq. \eqref{drfieldmisu}, where a differential operator is formally inverted -- so is $\chi^u$ in the dressed fields \eqref{dressed-A}-\eqref{dressed-F}} above. This indicates that the susy gauge symmetry is \emph{substantive} in MISU: its elimination ``costs" the field locality. This trade-off ``invariance \emph{vs} locality" is a typical signature of substantive gauge symmetries \cite{Francois2018}.
We also highlight that, since $u=u[\mathbb A]$, the susy-invariant dressed field $\mathbb A^{\boldsymbol{u}}$ -- and in particular $\psi^u=\gamma \chi^u$ -- is a  self-dressed \emph{relational} field variable. 

\smallskip
The dressed fields $\mathbb A^{\boldsymbol{u}}$ and $\mathbb F^{\,\boldsymbol{u}}$ are expected to exhibit residual $\SL(2, \CC)$-transformations. 
To find them, one need only determine the $\SL(2, \CC)$-transformations of the dressing: $\boldsymbol{u}^\mathbb B$.  
We have
\begin{equation}
\label{restr1stkinfmisu}
\begin{aligned}
u^B  \defeq&\, u[\mathbb A^B] 
= (b[\omega^B]^{-1})^{\,\mu} (\uprho^B_\mu) 
= B\-b[\omega]^{-1} B (B\- \uprho_\mu) 
= B\- b[\omega]^{-1} (\uprho_\mu) \\
=&\, B\- u[\mathbb{A}],
\end{aligned}
\end{equation}
where we used the fact that $b_
\mu[\omega]$ is a Lorentz covariant operator. 
This is cast in matrix form as
\begin{align}
\label{Lorentz-GT-u}
\boldsymbol{u}^{\mathbb{B}} 
 = \mathbb{B}\- \boldsymbol{u}\, \mathbb{B} 
  = 
\begin{pmatrix} B\- & 0\\ 0 & 1 \end{pmatrix}
\begin{pmatrix}
\1 & \  u[\mathbb A] \\
0 & 1
\end{pmatrix}
\begin{pmatrix} B & 0\\ 0 & 1 \end{pmatrix}
  =
\begin{pmatrix}
\1 & \  B\- u[\mathbb A] \\
0 & 1
\end{pmatrix}. 
\end{align}
Then, given that $\mathbb{A}^{\mathbb B}$ is given by \eqref{GT-A-F-MISU} (specialising $g\rarrow \mathbb{ B}$), it is immediate that   
\begin{equation}
\begin{aligned}
\label{B}
(\mathbb{A}^{\boldsymbol{u}})^{\mathbb{B}} &=  \mathbb{B}\- \mathbb{A}^{\boldsymbol{u}}\,  \mathbb{B} + \mathbb{B}\- \, d\,  \mathbb{B} =\begin{pmatrix}
     B\-\omega B + B\- \,dB &\ B\- \gamma \, \chi^u \\ 0 & 0
 \end{pmatrix}, \\[2mm]
(\mathbb{F}^{\,\boldsymbol{u}})^{\mathbb{B}}&= \mathbb{B}\- \mathbb{F}^{\,\boldsymbol{u}}\, \mathbb{B} = \begin{pmatrix}
     B\-\Omega B &\ B\- {(\gamma_a T^a \chi^u - 
 \gamma \, \nabla \chi^u )} \\ 0 & 0
 \end{pmatrix} .
\end{aligned}
\end{equation}
That is, the susy-invariant dressed field $\mathbb A^{\boldsymbol{u}}$ have standard residual Lorentz gauge transformations.

Up to now we have dealt with the  kinematics.
{As explained at the end of Section \ref{Matter-Interaction Supergeometric Unification via the DFM},
regarding the dynamics, one may consider two possibilities: 
Lagrangians that are (quasi-)invariant under either the initial full gauge supergroup $\H$, or  the residual gauge symmetries $\H/$susy $=\SL(2, \CC)$. 
The former case is more constraining, but then it is possible to dress such an $\H$-invariant Lagrangian. 
This was done, e.g., in \cite{JTF-Ravera2024ususyDFM} for the $3D$ AVZ model based on the superalgebra $\mathfrak{osp}(2|2)$, which is dressed following the prescription given by \eqref{pert-dressed-Lagrangian} --
see  Section \ref{The AVZ model as a MISU subcase} below. 
The latter case allows more freedom, as susy-invariance is kinematically guaranteed by the DFM (the basic dressed fields being susy singlets), and is arguably more sensible as it deals directly with the physical relational d.o.f. of the kinematics.}

\subsubsection{Dressed BRST formulation of the MISU model}
\label{Dressed BRST formulation of the MISU model}

The, naturally relational, MISU framework just described can also be recast in the BRST formulation, in which one can derive the dressed BRST algebra. Here we do this for the relevant four-dimensional example we just presented.

The BRST algebra of the simple model  above,  reproducing the linear gauge transformation \eqref{linear-GT-A-F}, is 
\begin{equation}
\begin{aligned}
    s \mathbb{A} = - D^\mathbb{A} \mathbb{C},
    \quad \text{ and } \quad 
    s \mathbb{F} = \left[\mathbb{F} , \mathbb{C}\right],
\end{aligned}
\end{equation}
where the ghost field splits as $\mathbb{C} = \CC_{\text{L}} + \CC_{\text{susy}}$ with a  Lorentz ghost and a susy ghost. In  matrix notation,
\begin{align}
    \mathbb{C} = \CC_{\text{L}} + \CC_{\text{susy}} = \begin{pmatrix}
    \beta & \  0 \\
    0 & 0
    \end{pmatrix} + \begin{pmatrix}
    0 & \  \varepsilon \\
    0 & 0
    \end{pmatrix} = \begin{pmatrix}
    \beta & \  \varepsilon \\
    0 & 0
    \end{pmatrix},
\end{align}
with  $\beta$ and $\varepsilon$ now being assigned ghost degree 1. 
The BRST operator splits accordingly as $s = s_{\text{L}} + s_{\text{susy}}$.
The BRST versions of the defining  susy transformation  \eqref{susy-GT-u} of the dressing  $\boldsymbol{u}[\mathbb A]$, and its $\SL(2, \CC)$ transformation \eqref{Lorentz-GT-u}, are 
\begin{align}
 s_{\text{susy}} \boldsymbol{u} = -\CC_{\text{susy}} \boldsymbol{u} 
     \quad \text{and} \quad 
 s_{\text{L}} \boldsymbol{u} = [\boldsymbol{u}, \CC_{_{\text{L}}}]. 
\end{align}
Thus, the dressed BRST algebra satisfied by $\mathbb{A}^{\boldsymbol{u}}$ and $\mathbb{F}^{\,\boldsymbol{u}}$ is 
\begin{equation}
\begin{aligned}
    s \mathbb{A}^{\boldsymbol{u}} = - D^{\mathbb{A}^{\boldsymbol{u}}} \mathbb{C}^{\boldsymbol{u}},
    \quad \text{and} \quad 
    s \mathbb{F}^{\,\boldsymbol{u}} = \left[\mathbb{F}^{\,\boldsymbol{u}} , \mathbb{C}^{\boldsymbol{u}} \right],
\end{aligned}
\end{equation}
 with  dressed ghost found to be
\begin{equation}
\begin{aligned}
    \CC^{\boldsymbol{u}} &= \boldsymbol{u}\- \CC \,\boldsymbol{u} + \boldsymbol{u}\- s \boldsymbol{u}\\
    & = \boldsymbol{u}\- (\CC_{\text{L}}+\CC_{\text{susy}}) \boldsymbol{u} + \boldsymbol{u}\- (s_\text{L} \boldsymbol{u} + s_{\text{susy}} \boldsymbol{u}) \\
    & = \boldsymbol{u}\- (\CC_{\text{L}} + \CC_{\text{susy}}) \boldsymbol{u} + \boldsymbol{u}\- ([\boldsymbol{u},\CC_{\text{L}}] - \CC_{\text{susy}} \boldsymbol{u}) \\
    & = \boldsymbol{u}\- \CC_{\text{L}} \boldsymbol{u} + \boldsymbol{u}\- (\boldsymbol{u} \CC_{\text{L}} - \CC_{\text{L}} \boldsymbol{u}) = \boldsymbol{u}\- \boldsymbol{u} \CC_{\text{L}} 
    = \CC_{\text{L}}.
\end{aligned}
\end{equation}
Which shows that the susy-invariant dressed fields $\mathbb{A}^{\boldsymbol{u}}$ and $\mathbb{F}^{\,\boldsymbol{u}}$ are standard Lorentz gauge fields, as we have
\begin{equation}
\begin{aligned}
    s \mathbb{A}^{\boldsymbol{u}} 
    =
    s_{\text{L}}\mathbb{A}^{\boldsymbol{u}}
    =
    - D^{\mathbb{A}^{\boldsymbol{u}}} \mathbb{C}_{\text{L}} = \begin{pmatrix}
    - \nabla \beta & \  - \gamma \,\beta \, \chi^u \\
    0 & 0
    \end{pmatrix} , 
    \quad \text{and} \quad 
    s \mathbb{F}^{\,\boldsymbol{u}}
     =
    s_{\text{L}}\mathbb{F}^{\,\boldsymbol{u}}
    =
     \left[\mathbb{F}^{\,\boldsymbol{u}} , \mathbb{C}_{\text{L}} \right] = \begin{pmatrix}
    \left[\Omega, \beta \right] & \  - \gamma \, \beta \,\nabla \chi^u \\
    0 & 0
    \end{pmatrix}.
\end{aligned}
\end{equation} 

This is just kinematics. Regarding the dynamics, let us again remark that one now only has to require a dressed Lagrangian $L^{\boldsymbol{u}}$ to be invariant under $s_{\text{L}}$; the dressed fields being in the kernel of $s_{\text{susy}}$, so will be a Lagrangian functional written for them.\footnote{We mention that this will be crucial to the closure of the standard supersymmetry algebra \emph{off-shell} on field variables, fundamental issue in standard supersymmetric field theory, but (dis)solved via the DFM, as we will show in the forthcoming paper \cite{JTF-Ravera2025offshellsusy}.}

\subsection{{The AVZ model as a MISU subcase}}\label{The AVZ model as a MISU subcase}

Let us close 
{by pointing out} that the MISU approach encompasses the so-called ``unconventional supersymmetry" introduced by Alvarez, Valenzuela, and Zanelli in the guise of a three-dimensional model
\cite{Alvarez:2011gd,Alvarez:2013tga}. 
In this ``AVZ model", based on the superalgebra $\mathfrak{osp}(2|2)$, a superconnection $\mathbb A_{\text{AVZ}}$ is chosen such that its susy part is \emph{required} to satisfy what has been dubbed the ``matter ansatz" in \cite{Alvarez:2021zhh}: $\psi\defeq \gamma\chi$. 
The field $\mathbb A_{\text{AVZ}}$ thus encodes both a gauge and a matter field, and the model has been shown to be applicable to the description of graphene systems, see e.g. \cite{Iorio:2014nda,Iorio:2018agc,Ciappina:2019qgj,Acquaviva:2022yiq}. 

The conceptual and technical status of the ``matter ansatz"  was unclear, and  \cite{Andrianopoli:2019sqe} 
attempted at interpreting it as a gauge-fixing \cite{Andrianopoli:2019sqe}.
It will not surprise the reader that 
{in} \cite{JTF-Ravera2024ususyDFM} 
{it was} showed to be better understood as a case of the DFM: 
In a way very similar as  above, a \emph{perturbative} susy dressing field $\upsilon[\mathbb{A}_\mathfrak{osp}]$ is extracted from the general $\mathfrak{osp}(2|2)$ superconnection $\mathbb{A}_\mathfrak{osp}$, again via  a gamma-trace decomposition of its odd component $\psi$. 
The perturbatively dressed connection then reproduces the AVZ ansatz with the benefit of achieving susy-invariance at 1st order, $\mathbb{A}_\mathfrak{osp}^\upsilon \simeq \mathbb A_{\text{AVZ}}$. 

The ususy literature struggled to extend it to higher dimensions: the main cause for puzzlement being the apparent necessity of ``two metric structures" (or vielbein), one from the manifold $M$ the other ``internal". See  e.g. \cite{Alvarez:2013tga, Alvarez:2020qmy, Alvarez:2021zhh, Alvarez:2023auf} for attempts at four-dimensional cases.
{The MISU approach}
decisively clarifie{s} this issue, {as the above example illustrates}, showing that difficulties stemmed from inadequate choices of superalgebras, and misunderstanding of the proper underlying Cartan supergeometric picture. 
Indeed, the issue was to {have chosen} 
superalgebras $\mathfrak g$ for the kinematics where, like the super-Poincaré algebra, the brackets of susy generators $\mathbb{Q}$ have a component along the translations (or transvections) generator $\mathbb{P}$.
This leaves with either of \emph{two options}: 
\begin{itemize}
    \item[i)] To treat the $\mathfrak g$-valued superconnection as an Ehresmann connection, acted upon by a gauge group $\mathcal G$. A move exactly analogous to \emph{Poincaré gauge gravity} or \emph{Metric-Affine Gravity} (MAG), and suffering all the same technical drawbacks and conceptual issues; chief among which the presence of unwanted ``internal gauge translations" and the presence of a ``translational gauge potential" somehow to be identified with (or made to induce) the vierbein on $M$.\footnote{{Let us mention, in echo to footnote 16, that in the Poincaré and Metric affine gravity (MAG) literature both the aforementioned  problems are ``solved" }by introducing \emph{ad hoc} the so-called ``radius vector" $\xi$ \cite{Trautman1973}, a $\mathbb{R}^n$-valued 0-form, which is none other than a \emph{dressing field for the translation gauge subgroup} of the full affine gauge group, leading to what is  called the ``key relation" of MAG in \cite{Hehl-et-al1995}. See \cite{JTF-Ravera-2025-MAG}.
    }
    \item[ii)] To consider a natural Cartan geometric picture for these algebras (if it exists): i.e., finding $\mathfrak{h}\subset \mathfrak{g}$ such that  $(\mathfrak{g}, \mathfrak{h})$ is a  super-Klein pair on which a Cartan supergeometry is modeled, $H$ being the fibration and the vielbein being $\mathfrak{g}/\mathfrak{h}$-valued. But this gives a superbundle over a \emph{supermanifold} rather than over a manifold, so that the fermionic 1-form field $\psi$ is part of a \emph{supervielbein} and susy is the odd part of the \emph{superdiffeomorphisms} of such supermanifold, rather than an internal gauge symmetry. This is, e.g., the case of ``standard" supergravity.
\end{itemize}
Unfortunately, none lead to a compelling unification of matter and gauge fields, contrary to the MISU approach sketched here, where it is achieved naturally.

\section{Conclusions and prospects of the novel setup}\label{Conclusion} 

We have shown that the DFM features decisively in simple, yet foundational, models of supersymmetric field theory.
In particular, we have shown that popular ``gauge-fixing conditions", when  solved explicitly,  yield dressing (super)fields.\footnote{Similarly, closer inspection reveals that many ``gauge-fixings" in the gauge field theory literature are actually instances of dressings: e.g. the Lorenz gauge, the harmonic gauge, the de Donder gauge, the axial gauge, or the so-called ``triangular gauge" \cite{Thiemann:2023lcy}. 
The latter being manifestly a special case of the Lorentz dressing field built for Cartan gravity in \cite{GaugeInvCompFields}, section 4.3., which also discuss its limitations. See also footnote 12 of \cite{Francois2018}.
This observation will be the object of a future work \cite{JTF-Ravera2025gfdressing}.}
This entails that one may build the RS spinor-vector field and the gravitino  as (\emph{non-local}) supersymmetry-invariant \emph{relational} dressed field variables, with the expected number of (off-shell, and then on-shell) degrees of freedom. 
The DFM  therefore  appears  foundational to susy field theory.

Yet, it bears emphasizing that some procedures/operations named ``dressing" in the susy literature are in no way related to the precise technical and conceptual meaning of the terminology as employed in the DFM. 
For example, in the context of (gauged) supergravity, in \cite{Inverso:2025zct} the object  $\mathbb{T}$, called $T$-tensor, is said to be the ``dressing of $\Theta$ [the embedding tensor] by the scalar field coset representative". Yet, this quantity is not a dressed field as defined here, since the ``dressing object" identified in eq. (8) of the aforementioned work is coset-valued, while a dressing field in the DFM is group-valued. 
The only way for the two notions to connect would be the specific case where the coset is a group, i.e. for the main group to be a direct or semi-direct product.

\medskip

Staying within standard supersymmetric field theory, a well-known issue is the a priori puzzling fact that the algebra of susy transformations close (in general) only on-shell.
In this regard, the fact that the DFM produces susy-invariant objects will reveal key to \emph{off-shell} susy \cite{JTF-Ravera2025offshellsusy}. 

As we observe in the introduction of Section \ref{Matter-Interaction Supergeometric Framework}, another issue of susy as usually conceived in high energy physics, is that it can only be saved from immediate empirical refutation by appealing to some mechanism of (spontaneous) susy breaking. 
In such more sophisticated susy models the DFM is bound to play a role too. 
Indeed, the reformulation of the electroweak (EW) model via the DFM showed that the notion of Spontaneous Symmetry Breaking (SSB) is superfluous to the empirical success of the theory: a gauge-invariant understanding of the  BEHGHK mechanism is available, preserving all the phenomenology and showing the physically operative mechanism to be a phase transition of the EW vacuum (without gauge symmetry ``breaking"). A conclusion reached, or hinted at, independently by many authors, among whom Higgs and Kibble themselves.
Indeed, while the DFM is a clean mathematical framework with a clear (relational) interpretive structure, the 
SSB is not an ``approach" to GFT but only an \emph{interpretation} of the BEHGHK mechanism -- a heuristic from the context of discovery of the EW model that has sedimented and  uncritically considered as part of its context of justification \cite{Francois2018}. 
The same will be shown to apply to ``susy SSB".

Relatedly, we observe that the DFM  also \emph{encompasses} the standard Stueckelberg trick to obtain the Proca model for a massive gauge field, the Stueckelberg field being a special case of dressing field -- not the other way around as mistakenly remarked in \cite{Grassi:2024vkb}. More broadly it underlies the so-called ``Massive Yang-Mills" models. 
Insofar, as Stueckelberg trick is used in susy models, the DFM shall bring further conceptual and technical clarity.
The DFM as a framework encompasses both the EW model and Stueckelberg-type models  \cite{JTF-Ravera2024gRGFT,Francois2018},
but ultimately, it is empirical adequacy that arbitrates between models. 

\medskip

The new MISU approach we introduced in Section \ref{Matter-Interaction Supergeometric Framework} does not suffer these problems plaguing standard susy:
it aims to exploit supergeometry and the DFM to unify  gauge and matter fields in a superconnection.
We illustrated the proposal with a simple 4D model base on a semi-direct super-extension of the Lorentz algebra. 
This approach encompasses the so-called ``unconventional supersymmetry" introduced in 3D, as shown in  \cite{JTF-Ravera2024ususyDFM}, and clarify why some authors struggled to extend it in four spacetime dimensions in a compelling way. 

To develop the proposal, one may work out more elaborated models than the one we presented. 
For example the class of models based on
orthosymplectic superalgebras  $\mathfrak{osp}$ containing $\mathfrak{spin}(1,3)\simeq\mathfrak{sl}(2, \CC)\simeq \mathfrak{sp}(2,\CC)$ in their even subalgebra -- such as the superalgebra $\mathfrak{osp}(2, \mathcal{N}; \CC)$ \cite{Abe-Nakanishi1988, Abe1989, ABE1990} -- fit our template for MISU. 
Another important class of scenarios fitting our template (in $d\geq 3$ in general) are super-Klein pairs $(\mathfrak{g}, \mathfrak{h})$ with $\mathfrak{h}\subset \mathfrak{g}$ such that $\mathfrak{g}/\mathfrak{h}$ is an even vector space, which lead to the desiderate Cartan supergeometric picture: a Cartan superbundle over a bosonic $M$ and a $\mathfrak g$-valued Cartan superconnection whose $\mathfrak{g}/\mathfrak{h}$-part is the vielbein $e^a$ of $M$. In this case, the dressed Cartan superconnection would display $e^a$ both in its $\mathfrak{g}/\mathfrak{h}$-part and in the odd sector of its $\mathfrak h$-part (analogously to what happens in the already discussed ``matter ansatz" of the AVZ model in $d=3$). 
This would supply the kinematics for Einstein-Dirac MISU models describing gravitational and matter fields unified in a single superconnection. 

{All these examples of application and open directions of the Matter-Interaction supergeometric framework we propose would strengthen its connection with pre-existing attempts by Berezin and collaborators to unify the description of bosonic gauge fields and fermionic matter fields, exploiting a single superconnection, without implying the existence of standard supersymmetry and independently of the phenomenological evidence of the latter.}

Furthermore, a very interesting parallel is to be drawn between the MISU approach  and antecedent proposals to use the framework of non-commutative geometry to describe gauge fields and the Higgs field as unified in a single non-commutative connection \cite{Dubois-Violette_kerner_Madore1990a,Dubois-Violette-Kerner-Madore1990b,Connes-Lott1991, Masson-Lazz2013,Masson-Lazz}.
Maybe the parallel is not so surprising, as supergeometry is in fact the ``mildest" form of non-commutative geometry. 
A fusion of both endeavours,  exploiting e.g. the setup of super-Lie algebroids together with the DFM, may lead to geometric unification of  gauge, matter and Higgs fields: all parts of one and the same generalised superconnection.
This investigation is left to future works.

\section*{Acknowledgments}  

We thank P. Berghofer for a careful reading of the manuscript and useful feedback.
J.F. is supported by the Austrian Science Fund (FWF), grant \mbox{[P 36542]}, 
and by the Czech Science Foundation (GAČR), grant GA24-10887S.
L.R. is supported by the 
GrIFOS research project, funded by the Ministry of University and Research (MUR, Ministero dell'Università e della Ricerca, Italy), PNRR Young Researchers funding program, MSCA Seal of Excellence (SoE), 
CUP E13C24003600006, ID SOE2024$\_$0000103, of which this paper is part.

\appendix

\section{Perturbatively dressed BRST formalism}\label{Perturbatively dressed BRST formalism}

Here we give the \emph{perturbatively dressed} BRST algebra, mentioned at the end of Section \ref{Dressed BRST algebra}. 
It is possible to proceed as follows: given 
the definition of the ``bare" BRST algebra \eqref{BRSTalgebradef}-\eqref{curvBRSTalg}  and of the perturbatively dressed fields
\eqref{pertdrfieldsgendef}-\eqref{pert-dressed-fields}, one shows that the latter satisfy a perturbative dressed BRST algebra,
\begin{equation}
\label{perturb-dressed-BRSTalgebradef}
\begin{aligned}
    & sA^\upsilon=-D^\upsilon c^\upsilon , \quad s \phi^\upsilon = -\rho_*(c^\upsilon)\,\phi^\upsilon, \\[2mm]
    & sF^\upsilon = [F^\upsilon,c^\upsilon] , \quad s(D^\upsilon\phi^\upsilon) = - \rho_*(c^\upsilon)\,D^\upsilon\phi^\upsilon, \\[2mm]
    & sc^\upsilon = - \frac{1}{2}[c^\upsilon,c^\upsilon].
\end{aligned}
\end{equation}
with the perturbatively \emph{dressed ghost} $c^\upsilon$ defined by
\begin{align}
    \label{dressed-ghostpert}
    c^\upsilon \defeq  c  + s \upsilon + [c, \upsilon].
\end{align}
This holds as a formal result irrespective of the BRST variation $s\upsilon$ of $\upsilon$, i.e. independent of  wether or not it is a dressing field. We stress if it obtain by neglecting terms of order 2 in $\upsilon$ in perturbation theory. 

Let us consider the case where $c=c_K+c_J$, and $s=s_K+s_J$, and where $\upsilon$ is a $\K$-dressing field: 
In that case its defining BRST transformation is 
\begin{equation}
    s_k \upsilon = -c_K,
\end{equation}
which is just the BRST version of \eqref{pert-dressing-field}.
We remark that such objects have appeared in the literature, without  being recognized as perturbative dressing field however.
Which is unfortunate given the technical and conceptual clarity brought by the DFM -- see e.g. appendix A of \cite{Grassi:2024vkb}, where the object $\Phi^K$ in formulas (A.6), (A.7), (A.8), (A.9) is actually a perturbative dressing field as defined in the DFM.
The perturbatively dressed ghost is then
\begin{equation}
    \label{perturb-dressed-ghost-bis}
\begin{aligned}
    c^\upsilon \defeq&\,  (c_K+c_J)  + s_K \upsilon+ s_J \upsilon + [c_k +c_J, \upsilon] \\
    =&\, c_J + s_J \upsilon + [c_J, \upsilon],
\end{aligned}
\end{equation}
where the term $[c_k, \upsilon] = -[s\upsilon, \upsilon]$ is to be neglected for consistency, as it is of order 2 in $\upsilon$ in perturbation theory.
Thus the perturbatively dressed BRST algebra \eqref{perturb-dressed-BRSTalgebradef}
manifestly encodes the residual $\J$-transformations of the perturbatively dressed fields. 
The BRST variation of the perturbatively dressed Lagrangian $L^\upsilon =L(A^\upsilon, \phi^\upsilon)$   is then 
\begin{align}
 s_K  L(A^\upsilon, \phi^\upsilon) =0,
 \quad \text{ and } \quad
  s_J L(A^\upsilon, \phi^\upsilon) = d\beta(A^\upsilon,\phi^\upsilon;c^\upsilon),
\end{align}
with $c^\upsilon$ given by \eqref{perturb-dressed-ghost-bis}. This shows that after $\K$-dressing, the dressed Lagrangian is in the kernel of $s_K$, but belongs to the $s_J$ modulo $d$ cohomology, in ghost degree $0$: $L^\upsilon \in H^{0|n}(s_J|d)$.

In the particular case where  the residual BRST variation of the  dressing field $\upsilon$ is 
\begin{equation}
    s_J \upsilon = [\upsilon, c_J],
\end{equation}
which is just the BRST version of the linear residual transformation of the dressing mentioned in footnote 7, we have that the residual perturbatively dressed ghost is
\begin{equation}
\begin{aligned}
    \label{perturb-dressed-ghost-bis-bis}
    c^\upsilon 
    &= c_J + s_J \upsilon + [c_J, \upsilon] \\
    &= c_J + [\upsilon, c_J ] + [c_J, \upsilon] = c_J.
\end{aligned}
\end{equation}
Which means that the perturbatively dressed fields are standard $\J$-gauge fields, satisfying a standard residual (perturbative) BRST algebra, given by \eqref{perturb-dressed-BRSTalgebradef} with dressed ghost $c^\upsilon=c_J$. 
The residual BRST variation of the perturbatively dressed Lagrangian $L^\upsilon$  is then 
$s_J L(A^\upsilon,\phi^\upsilon)=d\beta(A^\upsilon,\phi^\upsilon;c_J)$, showing it to be a standard $\J$-gauge theory.

{
\normalsize 
 \bibliography{Biblio11}
}

\end{document}